\documentclass[%
aps,
nofootinbib,
preprint,%
]{revtex4-2}

\usepackage{graphicx}
\usepackage{dcolumn}
\usepackage{bm}

\usepackage{etoolbox}
\usepackage{url}
\usepackage{hyperref}
\usepackage{amsmath,amsthm, amssymb} 
\usepackage{enumerate} 
\usepackage{float, subfig} 
\usepackage{color} 
\usepackage{xspace} 
\usepackage{braket} 
\usepackage{relsize}
\usepackage[nodisplayskipstretch]{setspace}
\usepackage{array}
\newcolumntype{P}[1]{>{\centering\arraybackslash}p{#1}}
\makeatletter
\def\@email#1#2{%
 \endgroup
 \patchcmd{\titleblock@produce}
  {\frontmatter@RRAPformat}
  {\frontmatter@RRAPformat{\produce@RRAP{*#1\href{mailto:#2}{#2}}}\frontmatter@RRAPformat}
  {}{}
}%
\makeatother

\renewcommand{\bold}[1]{\boldsymbol{#1}} 

\renewcommand{\v}[1]{\ensuremath{\mathbf{#1}}} 
\newcommand{\gv}[1]{\ensuremath{\mbox{\boldmath$ #1 $}}}
\newcommand{\avg}[1]{\left< #1 \right>} 
\newcommand{\pd}[2]{\frac{\partial #1}{\partial #2}} 
\newcommand{\grad}[1]{\gv{\nabla} #1} 

\newcommand{\mode}{\!\left(\substack{\gv{k} \\ \nu}\right)}

\newcommand {\eqn}[1] {Eq.~(\ref{#1})}
\newcommand {\eqns}[1] {Eqs.~(\ref{#1})}
\newcommand {\eqnx}[1] {(\ref{#1})}

\newcommand {\sect}[1] {Section~\ref{#1}}

\newcommand {\fig}[1] {Fig.~\ref{#1}}

\newcommand {\figs}[1] {Figs.~\ref{#1}}
\newcommand {\figx}[1] {\ref{#1}}

\newcommand{\sumdd}[3]{\ensuremath{\sum_{\substack{#1 \\ #2}}^{#3}}}

\newcommand{\deltmd}{\Delta t_{\rm MD}}

\begin{document}


\title[]{Decoupling the effects of ripples from tensile strain on the thermal conductivity of graphene and understanding the role of curvature on the thermal conductivity of graphene with grain boundaries}
\author{Abhishek Kumar}
\author{Kunwar Abhikeern}
\author{Amit Singh}%
\email{amit.k.singh@iitb.ac.in}
\thanks{Corresponding author}
\affiliation{ 
Department of Mechanical Engineering, IIT Bombay, Mumbai 400076, India
}%

\date{\today}

\begin{abstract}
Ripples, curvature, and grain boundaries in graphene can significantly alter its thermal conductivity (TC), which is paramount in various applications including nanoelectronics and thermal management. In this study, we conducted extensive equilibrium molecular dynamics simulations and used the Green-Kubo method to elucidate the impact of ripples on the TC of graphene and the impact of curvature and tilt angles characterizing a grain boundary (GB) on the TC of polycrystalline graphene. Although tensile and compressive strains have been known to control the amount of ripples, the effects of strain and ripple on the TC have not been decoupled. With the
help of Green-Kubo simulations on larger graphene samples of size $150 \times 100$~\AA$^2$ and simulations based on the spectral energy density method on smaller samples without ripples of size $31 \times 26~ $\AA$^2$~, we show that both samples show an approximately
30\% decrease in TC between tensile strains 3\% and 10\% when ripples become negligible even in the larger sample. Between 0 and 3\% strain, when both ripples and strains are present in the larger sample, we decouple the effect of ripples from strain on the TC and show that ripples alone reduce the TC of the larger unstrained sample by approximately 61\%. We also present a unique technique
to introduce curvatures in the graphene sheet containing GBs by performing MD simulations under NPT ensembles with a constant pressure of 200 bar along the x-axis. 
Our analysis suggests that the Green-Kubo TC linearly decreases with curvature; however, the rate of decrease depends on the tilt angles of the grain boundary. 
An analysis of the TC in graphene samples with GBs of varying tilt angles reveals that the TC strongly depends on the tilt angles.
In summary, our research underscores the pivotal role of ripple, curvature, and grain boundary in modulating the thermal conductivity of graphene with and without the grain boundary. 
\end{abstract}

\maketitle
\section{Introduction}\label{sec:introduction}
Since the discovery of graphene~\cite{novoselov_2005},  modulating the ripples in graphene has been of paramount importance for various applications. However, the effect of ripples on the thermal conductivity (TC) of single-layer graphene (SLG) has not been well understood. 
The existence of static elastic ripples (corrugations) in graphene sheets was first confirmed by
Meyer et al.~\cite{meyer_2007} in 2007; however, Bangert et al.~\cite{bangert_2009} found that the patterns of ripples are not static but change with time.  With the help of atomistic Monte Carlo simulations, Fasolino et al.~\cite{fasolino_2007} attributed the spontaneous appearance of ripples to thermal fluctuations. These initial attempts to characterize the ripples in single-layer graphene soon led to manipulation of the sinusoidal ripples with strain-based graphene engineering~\cite{bao_2009}.
Molecular dynamics (MD) simulations were also performed to show that
the amplitude and orientation of the ripples can be controlled by different plane strain components~\cite{baimova_2012} or under the simultaneous action of shear and tensile membrane forces~\cite{dmitriev_2012}. The effect of strain on the TC of different graphene structures with associated changes in the amplitudes of the ripples has also been studied.
With non-equilibrium MD (NEMD) simulations, 
Zhang et al.~\cite{zhang_2013} did not find changes in the TC of 16 nm x 6 nm graphene nanoribbons (GNR) under 0 to 7.5\% tensile strain; but, they observed a 20-30\% decrease 
in the TC with 9-15\% tensile strain, which also coincides with the transformation of the ripples into structural ridges.  MD simulations of silicon doping in graphene sheets revealed a reduction in TC with irregular twisting of the smooth ripples due to the strain induced by doping~\cite{lee_2015}. With the help of
approach-to-equilibrium molecular dynamics simulations, Hahn et al.~\cite{hahn_2016} also found a reduction in TC
of polycrystalline graphene for grain sizes greater than 5~nm and suggested increased
out-of-plane corrugation as the possible reason behind the reduction.
Park et al.~\cite{park_2017} were able to generate strong ripples by applying compressive strains of 5 to 40\% along both in-plane orthogonal directions in the flat graphene sheet and
significantly reduced the TC. It can be concluded from these studies that a strain-induced increase in ripples reduces the TC of a flat graphene sheet; however, MD-based calculations have shown
only the combined effects of both strains and ripples on the TC. So far, there have been no attempts to decouple the effect of strains and ripples on the TC of graphene sheets.
In the present work, we attempt to fill this gap by isolating the individual
effects of strains and ripples using equilibrium MD simulations on pristine single-layer graphene (SLG). We also study the
effect of strain on the phonon properties, such as the group velocities and phonon
lifetimes. The phonon group velocities are calculated with the help of the harmonic lattice dynamics based GULP package and phonon lifetimes are obtained 
with the help of the robust and consistent normal mode decomposition (NMD) analysis based SED method~\cite{abhikeern2023consistent,mcgaughey2014predicting}.

Curvature is another important aspect through which the thermal conductivity of materials
can be modulated. Using Monte Carlo-based simulations of phonon transport,
Liang-Chun Liu et al.~\cite{liu_2009} evaluated the effect of curvature on TC of
silicon nanowires and were able to show a reduction of around 15\% in
TC of curved wire compared to straight wire. 
For silicon nanowires, using NEMD simulations, Xiangjun Liu et al.~\cite{liu_2019} have
also reported a $10 \%$ reduction in TC with an increase in curvature.
Mortazavi et al.~\cite{mortazavi_2012} performed NEMD simulations on the geometrically
prepared curved GNR and found that around 25\% reduction in TC can be
achieved when the temperature gradient develops along the circumferential direction in
the GNR with 350$^\circ$ circular arc angle.
Li et al.~\cite{li_2013} studied the effect of curvature on the ballistic TC of GNR using the non-equilibrium Green Function approach and reported an almost negligible reduction in TC when the heat current develops along the longitudinal direction.
Zhang et al.~\cite{zhang_2017} studied the effect of Gaussian curvatures on thermal conduction in carbon crystal with the help of Green-Kubo atomistic simulations and reported several order reductions in the TC when curving from zero to negative curvature structure.  
In the same year with the help of reverse NEMD simulations, Zhuang et al.~\cite{zhuang_2019} reported a reduction in TC of more than $90 \%$ along the bending direction and a reduction of around $25\%$ in TC perpendicular to the bending direction. 
These studies have been performed on graphene sheets without grain boundaries (GBs), where
the curvatures have been introduced geometrically. We, for the first time, study the
effect of curvature on the TCs of the graphene sheets with GBs. 
In our study, we introduce a unique methodology for introducing curvature in graphene sheets with GBs by performing equilibrium MD simulations under NPT ensemble conditions. 

We also study the effect of symmetric tilt GBs on the TCs of the graphene sheet using Green-Kubo simulations. Abhikeern and Singh ~\cite{abhikeern2024latticethermalconductivityphonon} provided a detailed review of the effect of GB on bicrystallline and polycrystalline
graphene samples.
Fox et al.~\cite{fox2019thermal} performed NEMD simulations on bicrystalline
graphene nanoribbons (bi-GNR) for a wide range of tilt angles under arbitrary in-plane 
thermal loading directions, and showed that the TCs decrease from 0$^\circ$ to 32.2$^\circ$ 
tilt angles and then increase almost symmetrically to 60$^\circ$ tilt angles. This
dependence on tilt angles is similar to the experimental work by Lee et al.~\cite{lee2017dependence} for bicrystalline graphene samples. Similarly, we also
show the dependence of TC on the GB tilt angles from 0$^\circ$ to 32.2$^\circ$ for a
graphene sheet of size $150 \times 100$~\AA$^2$, however, we used Green-Kubo simulations
instead of NEMD.  Green-Kubo simulations on polycrystalline graphene samples with an
average grain size from 1~nm to 5~nm had been performed by 
Mortazavi et al~\cite{mortazavi2014}. On the other hand, Liu et al~\cite{liu2014grain} 
also used the Green-Kubo method to show that TC decreases exponentially with
increasing GB energy for polycrystalline samples. However, they prepared the polycrystalline
samples with random tilt angles and did not study the dependence of TC on the GB tilt angles.
We perform Green-Kubo simulations over the graphene sheet with a single GB for the first time, though the polycrystalline effect kicks in because of the periodic boundary conditions imposed along the in-plane x and y axes. Similarly to Fox et al.~\cite{fox2019thermal}, we also show that the TC strongly depends on the tilt angles. 

The structure of the paper is as follows. Section 2 describes the preparation of the sample and Section 3 describes the
methodology for the Green-Kubo method, the SED method, and other simulation details. In Section 4, we investigate the effects of ripples and strains on the TC of pristine graphene samples. In Section 5, we discuss the effect of curvature on TCs of graphene samples with
two different GB tilt angles. Section 6 deals with the effect of different grain boundaries with different tilt angles on the TC. 
Finally, in Section 7 we conclude with a summary and suggestions for future work.
\section{Sample preparation}~\label{sec:sample_preparation}
\begin{figure}[htp]
    \centering
         \includegraphics[width=0.7\textwidth]{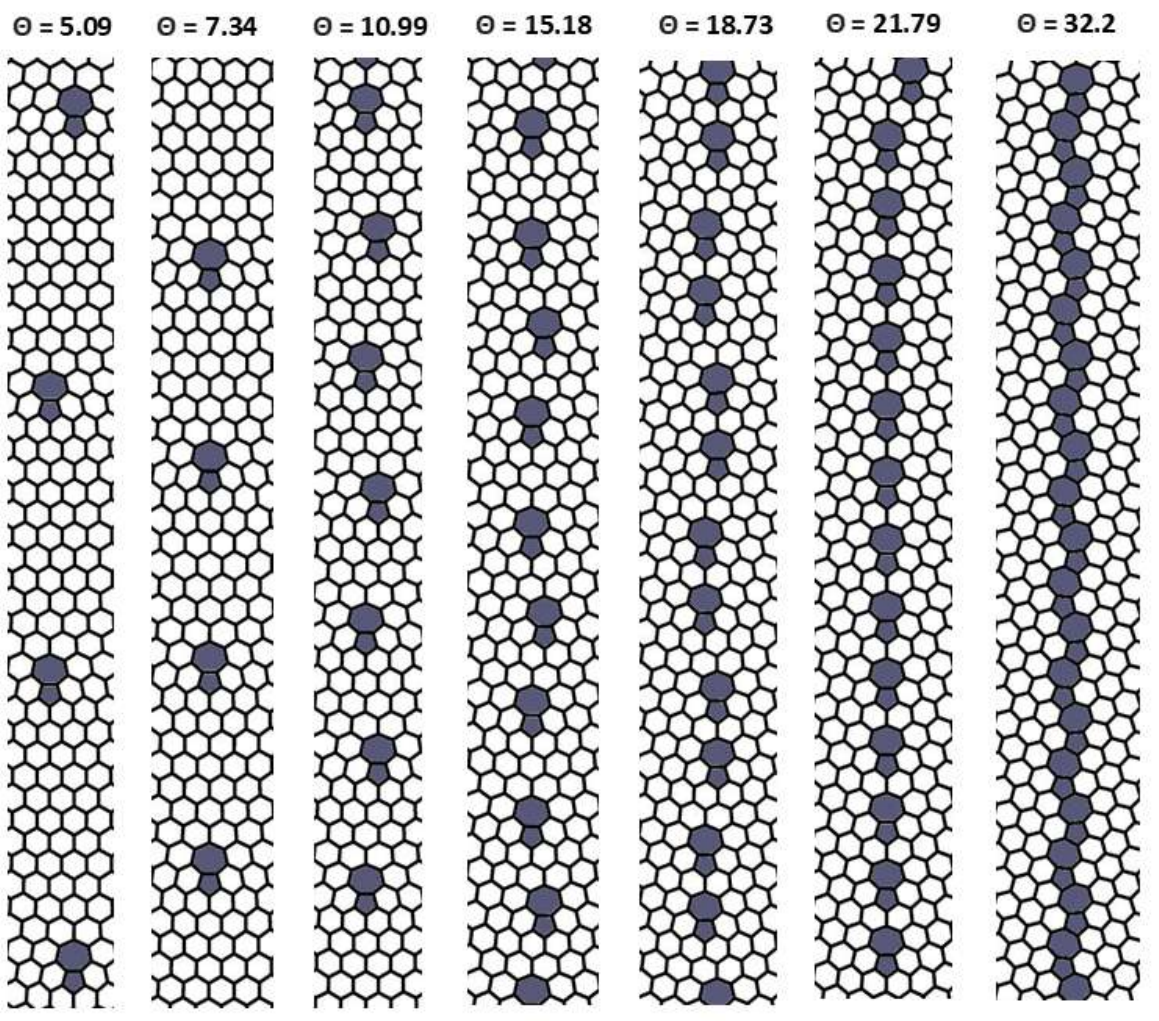}
     \caption{Graphene grain boundaries for seven different tilt angles $\theta =
     5.09, 7.34, 10.99, 15.18, 18.73, 21.79, 32.2$ degrees.}
     \label{fig01}
\end{figure}

Pristine graphene samples without grain boundaries are prepared so that the zigzag edges are along the x-axis and the armchair edges are along the y-axis. 
Samples with symmetric grain boundaries are prepared with the centroidal Voronoi tessellation (CVT) based algorithm proposed by Ophus et al.~\cite{ophus2015large}. 
Seven different tilt angles were selected between 0$^\circ$ and 32.2$^\circ$, where each tilt angle $\theta = \theta_L + \theta_R$ can be defined using translation vectors $(n_L, m_L)$ and $(n_R, m_R)$ corresponding to the left and right domains along the GB, respectively~\cite{zhang2013structures}.
 For symmetric GBs, we have $n_L=n_R=n$, $m_L=m_R=m$,
 $\theta_L = \theta_R = \frac{\pi}{3} - \arctan\frac{(2n + m)}{\sqrt{3}}$ and $n$ and $m$ are integers~\cite{ophus2015large, zhang2013structures}.
These grain boundaries have been shown in \fig{fig01} for seven different tilt angles.

\begin{figure}[htb!]
\centering
\includegraphics[width=0.5\textwidth]{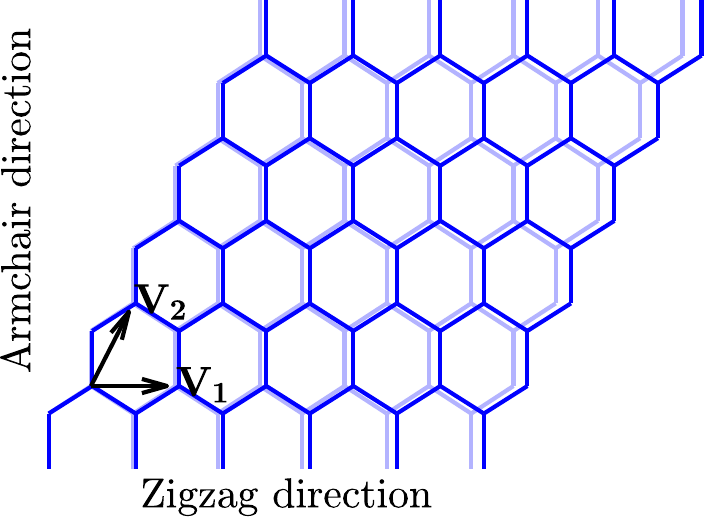}
  \caption{Comparison of unstrained (light blue) and $3 \%$ strained pristine graphene (dark blue). Strain is applied along the zigzag direction.}
  \label{fig02}
\end{figure}

When we study the effect of tensile strain on the pristine graphene with the help of the SED method, 
the two planar lattice vectors of the primitive unit cell of the SLG samples are taken as $\v{V}_1 = a(1+\rm{\epsilon_{zig}}) \hat{\v{i}}$ and $\v{V}_2 =  a((1+\rm{\epsilon_{zig}})\cos 60^\circ \hat{\v{i}} + \sin 60^\circ \hat{\v{j}})$, where $\hat{\v{i}}$ and $\hat{\v{j}}$ are unit vectors along the axes x and y, respectively, as shown in \fig{fig02}, and  $a = 2.492$~\AA~.
The term $\rm{\epsilon_{zig}}$ represents the amount of tensile strain along the x axis.
In \fig{fig02}, the unstrained and the strained SLG's have been shown in light blue and dark blue colors, respectively.  
\section{Computational Details}\label{sec:comp_details}
\subsection{Green-Kubo method}\label{sec:greenkubo}
We study the effects of ripples, curvature and grain boundaries on the thermal
transport in graphene with the Green-Kubo method, which gives the following relation
for the thermal conductivity tensor~\cite{evans:2008}:
 \begin{align}
   \bold{\kappa} = \frac{1}{V k_{\rm B} T^2} \int_{0}^{\infty}
   \avg{\bold{J}(t) \otimes \bold{J}(0)}_{\grad{T} = 0}\,dt,
   \label{GKeqb}
 \end{align}
 where $V$ is the volume of the system and $\bold{J}$
 is the microscopic heat current vector. The thermal conductivity is calculated at temperature $T$ and $\avg{\bold{J}(t) \otimes \bold{J}(0)}_{\grad{T} = 0}$ 
 represents the ensemble average of the heat current autocorrelation tensor function 
 for equilibrium ensembles without any temperature gradient, that is, $\grad{T} = 0$. To avoid infinity in the upper limit of the integration, for equilibrium MD based
 calculations, the thermal conductivity tensor in \eqn{GKeqb}
 can also be practically written as~\cite{singh:2015}
 \begin{align}
  \begin{split}
   \bold{k} &= \lim_{\tau_I \to \infty} \, \frac{1}{V k_{\rm B}
   \theta^2}\int_{0}^{\tau_I}
   \avg{\bold{J}(t) \otimes \bold{J}(0)}_{\grad{T} = 0}\,dt\\
   &= \lim_{\substack{S,I \to \infty\\S \gg I}} \frac{\deltmd}{V k_{\rm B}
   \theta^2}  \sum_{a = 1}^{I} \frac{1}{S-a} \sum_{b = 1}^{S-a}
   \bold{J}_{a+b} \otimes \bold{J}_b,
   \label{discreteGKeqb}
 \end{split}
 \end{align}
 where $\deltmd$ is the MD time step, $S$ is the total number of simulation steps,
 $I$ is the total number of integration steps, $\tau_I$ is the integration
 time ($\tau_I = I \deltmd$), and $\bold{J}_m = \bold{J}(m\deltmd)$.
 For the calculation of the heat current vector $\bold{J}$, we consider a system of 
 $N$ atoms with $m_i$ as the mass of the $i^{\rm{th}}$ atom, $\bold{r}_i$ as its
 position vector and $\tilde{\v{v}}_i$ as its velocity relative to the velocity of the center-of-mass of the system.
 The heat current $\bold{J}$ for systems with pair potential interactions can be written as~\cite{singh:2015,irving:1950}:
 \begin{align}
   \bold{J} = \sum_{i = 1}^{N} e_i \tilde{\v{v}}_i +
              \sumdd{i,j}{i<j}{N} \left[ \bold{f}_{ij}\cdot
                         \frac{\tilde{\v{v}}_i+\tilde{\v{v}}_j}{2} \right]
                         (\bold{r}_i - \bold{r}_j).
   \label{microCurrent}
 \end{align}
 Here, $e_i$ represents the local site energy of atom $i$ which includes both
the kinetic and the potential energy terms and $\bold{f}_{ij}$ is the force on atom $i$ due to atom $j$~\cite{singh:2015}. This definition, which is obtained
by the Irving-Kirkwood procedure~\cite{irving:1950}, can
be extended to interatomic potentials with three-body terms. 

All simulations are performed with LAMMPS~\cite{lammps} software. 
For all of our Green-Kubo based simulations, we first perform energy minimization with periodic boundary conditions (PBC) along the x and y axes and a free boundary condition was employed along the z axis. C-C interactions are modeled by the optimized Tersoff potential~\cite{Lindsay2010}.
The initial velocities of all the atoms are randomly selected from a Maxwell-Boltzmann distribution corresponding to twice the target room temperature so that
after equipartition of energy, the final temperature quickly equilibrates to the room temperature. The velocities are also adjusted
to ensure that the system has zero linear and angular momentum.
The equilibrium MD simulation is then performed with multiple MD runs where the MD time step is
 taken as $\deltmd=1$~fs. In the first run, the atoms were simulated under the NPT ensemble conditions with zero pressure and 300~K temperature for $10^6$ MD steps. For the next run, the NVT ensemble conditions are used at 300~K temperature for another $10^6$ MD steps. 
 The Nos\'{e}-Hoover chain
 thermostats are employed for these NPT and NVT runs. The final run is performed for another $6 \times 10^{6}$
 MD time steps under NVT conditions to ensure steady-state conditions. 
 We created five replicas of this NVT ensemble by changing the initial velocities of the atoms from the Maxwell-Boltzmann distribution and collected heat current vectors for this last run for each time step for each replica to perform a Green-Kubo analysis. The components of the thermal conductivity tensor $\kappa_{\rm{xx}}$ and $\kappa_{\rm{yy}}$ as functions of integration time $\tau_I$ are calculated using \eqn{discreteGKeqb} for each replica. The values of the TC components oscillate around a plateau within a range of $\tau_I$, whose plateau average within this range is computed. We collect these averaged values for all five replicas, and finally the mean and standard error of these five data are reported each for $\kappa_{\rm{xx}}$ and $\kappa_{\rm{yy}}$.
 The thermal conductivity $\kappa$ of the system is defined as the average of the in-plane diagonal components of the thermal conductivity tensor, that is, $\kappa = \frac{1}{2} (\kappa_{\rm{xx}} + \kappa_{\rm{yy}})$.

\subsection{SED method}\label{sec:SED}
For the analysis of the effect of strain on the thermal transport and phonon lifetimes, we utilize the SED method  which has been explained in details in Abhikeern and Singh~\cite{abhikeern2023consistent}.
 Based on the phonon Boltzmann transport equation (BTE) under relaxation time approximation and the Fourier's law,  the TC tensor of a system is given   by
  \begin{align}
      \bold{\kappa} = \sum_{\substack{\gv{k}}} \sum_{\substack{\nu}} c_{v}\mode \left(\v{v}_g\mode \otimes \v{v}_g\mode\right)\tau\mode.
      \label{kSED}
  \end{align}
  Here, the summation is over all allowed normal modes in the first Brillouin zone (BZ),
  and each mode $\mode$ is denoted by wavevector $\gv{k}$ and dispersion branch $\nu$. Moreover, $c_{v}\mode$,  $\v{v}_g\mode$ and $\tau\mode$   are the mode specific volumetric specific heat, group velocity and phonon lifetime, respectively.   We take $c_{v}\mode 
  =k_B/V$~\cite{mcgaughey2014predicting} for all modes for classical MD simulations, where $k_B$
  is the Boltzmann's constant and $V$ is the volume of the simulation box.
  The group velocity $\v{v}_g\mode =\pd{\omega\mode}{\gv{\kappa}}$ is obtained by using finite
  difference method over a fine grid  of wavevectors around the given mode $\mode$. The  GULP
  package~\cite{GULP} is used for calculating the normal mode frequencies $\omega\mode$ and
  the corresponding eigenvectors. We use GULP to also obtain the phonon density of states (PDOS).
  The phonon lifetime for a mode is calculated by fitting a Lorentzian function over
  the SED curve, which is determined by the normal mode decomposition (NMD) analysis by projecting the
  equilibrium MD simulation based atomic positions and velocities onto the normal mode
  coordinates~\cite{abhikeern2023consistent}. Like Green-Kubo method, we report
  the final TC as the average of the in-plane diagonal components of the thermal conductivity tensor.
 
 For all our SED-based simulations, the energy minimization is followed by an equilibrium MD simulation consisting of multiple MD runs with a time step of $\deltmd = 0.2$~fs. The first run is conducted under NVT conditions at a temperature of 300~K for $5 \times 10^5$ MD steps utilizing the Nos\'{e}-Hoover chain thermostat. The second run is performed under NVE conditions for another $10^5$ MD steps. Subsequently, an additional $2^{16}$ MD steps are performed under NVE conditions to ensure that steady-state conditions are reached without the influence of thermostats. Atomic positions and velocities were recorded for this last MD run every $2^2$ time step.
 Furthermore, following the approach of Qiu and Ruan~\cite{qiu2012molecular}, we use both the symmetry of the hexagonal Brillouin zone (BZ) and the ease of discretization to select allowed wavevectors within the first quadrant of the BZ, which have symmetric counterparts in the other quadrants. SEDs are calculated for all wavevectors allowed corresponding to the first BZ. These are then averaged over five MD runs with different initial velocities and further averaged over for each of the chosen symmetric wavevectors in the first quadrant before performing the Lorentzian fitting of these SED data to determine the phonon properties.

\section{Effect of ripples and strain on thermal conductivity}\label{sec:ripple}
In this section, our focus is on exploring the impact of ripple on thermal conductivity.  
We characterize the ripples in terms of their magnitude ($\Delta Z$) which we quantify as the time average of the absolute difference between the maximum and the minimum out-of-plane (z) coordinates of the sample, that is,
\begin{align}
  \begin{split}
     \Delta Z &= \lim_{\tau \to \infty} \, \frac{1}{\tau} \int_0^\tau \left|z_{\rm{max}}(t)-z_{\rm{min}}(t)\right|\,dt,\\
     &=\lim_{I \to \infty} \frac{1}{I}  \sum_{a = 1}^{I} \left|z_{\rm{max}}(a\deltmd)-z_{\rm{min}}(a\deltmd)\right|,
     \label{ripple}
  \end{split}
\end{align}
where we take $I = 6 \times 10^{6}$ MD steps with $\deltmd = 1$~fs as stated in \sect{sec:greenkubo}. We first consider a pristine graphene sample of size $150 \times 100$~\AA$^2$ with 150~\AA~length along x-axis and 100~\AA~ width along y-axis,
and calculate its thermal conductivity with the help of \eqns{discreteGKeqb} and \eqnx{microCurrent}. The thickness of the SLG is taken as  3.34~\AA, which is the interlayer spacing between two layers.
We finally obtain a value of $785 \pm 60$~W/mK for the TC of this sample
following the simulation details suggested in the Green-Kubo framework in \sect{sec:greenkubo}.
\begin{figure}[htp]
\centering
\includegraphics[width=1.0\textwidth]{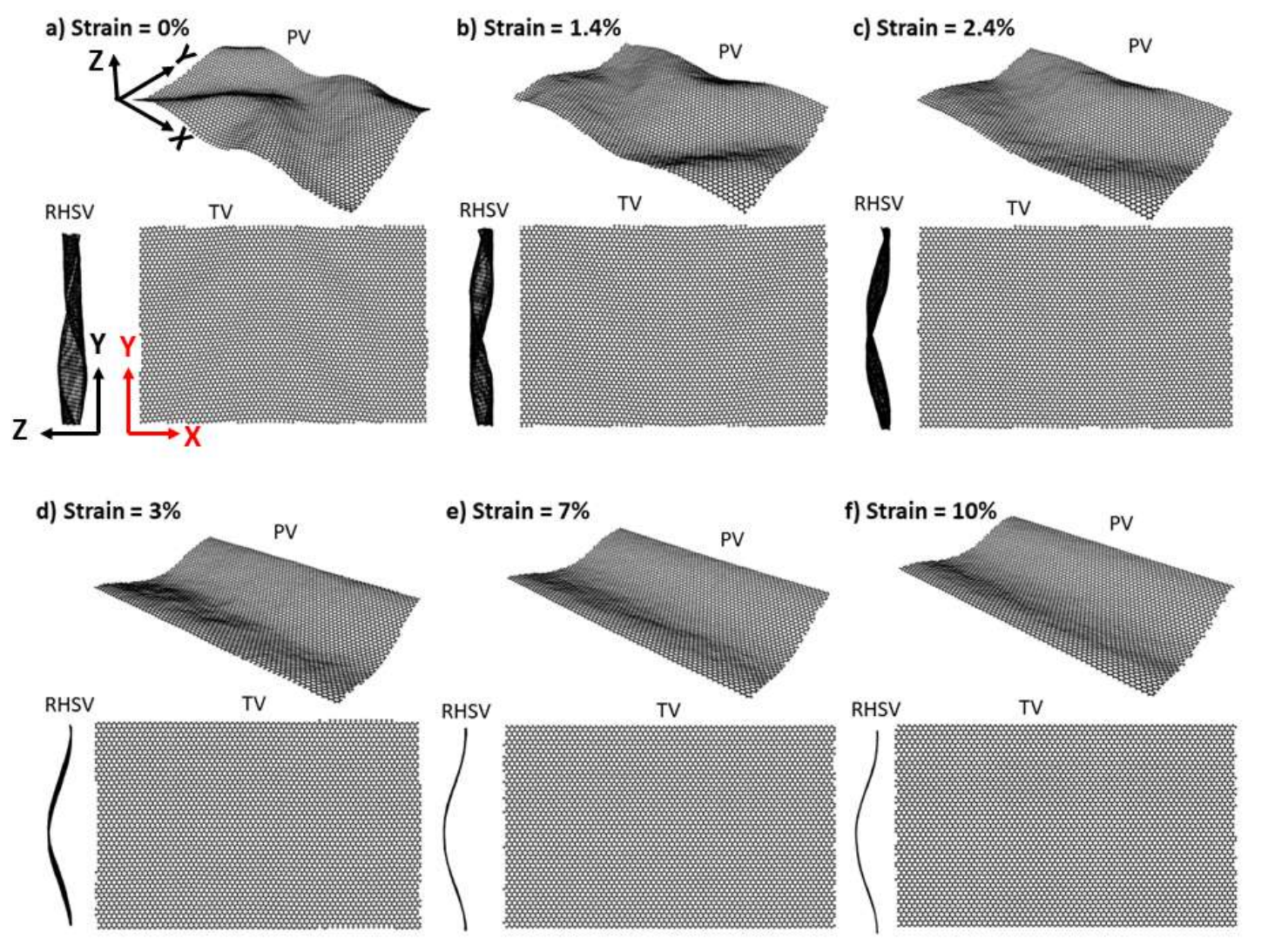}
  \caption{Structural changes in pristine graphene with the application of tensile strain along x axis. Beyond 3\% strain, ripples are almost negligible. PV, TV and RHSV refer to perspective view, top view and righ-hand side view, respectively.}
  \label{fig03}
\end{figure}
 This sample contains intrinsic ripples that can be observed in \fig{fig03}(a), whose magnitude calculated through \eqn{ripple} is 14~\AA. Now, it turns out, applying tensile strain along the x axis decreases the magnitude of ripples present in the system as shown in \fig{fig03}. We calculate the Green-Kubo TC and the magnitude of ripple $\Delta Z$ of the pristine graphene of size $150 \times 100$~\AA$^2$ under different tensile strains and show the results in \fig{fig04a} and \fig{fig04b}, respectively. 
\begin{figure}[htp]
    \centering
    \subfloat[$\kappa$]{\includegraphics[width=0.45\textwidth,height=2.2in]{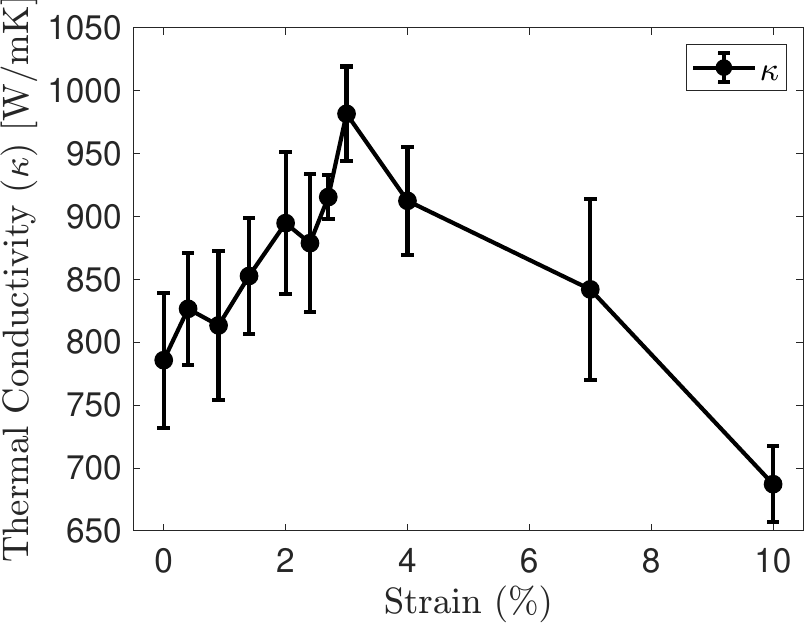}\label{fig04a}}
    \subfloat[$\Delta Z$]{\includegraphics[width=0.45\textwidth,height=2.2in]{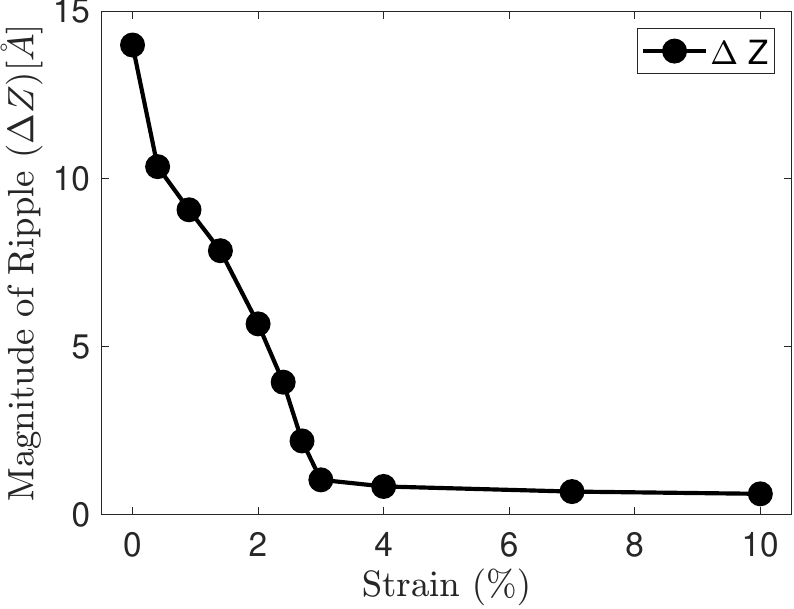}\label{fig04b}}
    \caption{Tensile strain dependence of (a) thermal conductivity, $\kappa$, and (b) magnitude of ripple, $\Delta Z$, calculated with the help of \eqn{ripple} for pristine graphene of size $150 \times 100$~\AA$^2$ .}
    \label{fig04}
\end{figure}
It can be observed both from \figs{fig03} and from \figx{fig04b} that the ripples are almost
absent for a tensile strain greater than 3\%\footnote{Beyond 3\% strain we do not observe ripples even in SLG samples with grain boundaries.}. This is also evident from \fig{fig04a} that the TC continuously decreases as the strain increases beyond 3\%. Therefore, it can
be said without any doubt that only the tensile strain is responsible for the decrease in TC beyond the strain 3\%. Between 3 and 10\% strain, we observe a decrease of approximately 30\% value in the TC. However, the same cannot be said for the strain range 0-3\%,
where both ripples and strain contribute to the TC. In fact, TC has increased by
approximately 25\% between 0 and 3\% strain. In this 0-3\% strain range, the effect of strain on TC should have followed the same secular trend as it followed in the 3-10\% strain range,
that is, TC should have decreased had only the strain effect been present. Therefore, the increase in TC must be attributed to the reduction in ripples. 
In the following, we perform a strain analysis using the SED method to better understand the effect of strain between 0 and 3\% and between 3 and 10\%.

\subsection{Quantifying strain effect on TC}\label{sec:strain}
\begin{figure}[htp]
\centering
\includegraphics[width=0.5\textwidth]{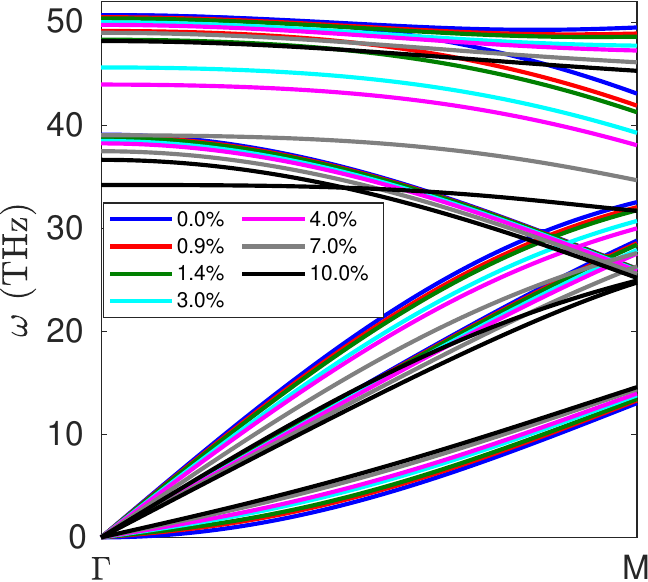}
\caption{Dispersion curve comparison of unstrained and strained graphene along $\Gamma-\rm{M}$ direction.}
  \label{fig05}
\end{figure}

We select $N_1 \times N_2$ primitive unit cell samples, where $N_1 = N_2 = 12$ which corresponds to a sample size of $31 \times 26~ $\AA$^2$~, to understand the effect of strain on the TC. We adopt the simulation strategy as mentioned in \sect{sec:SED}. 
With the help of \eqn{ripple}, we quantify the magnitude of ripple $\Delta Z = 1 $~\AA~. 
For all practical purposes, we can assume that this is almost negligible, and therefore
we can state that these small samples are almost ripple-free. 
In previous studies also, it has been shown that as we continue to increase the size of a graphene sheet the ripples begin to become more pronounced~\cite{deng2016wrinkled}, and therefore smaller sample sizes will have negligible ripples.
We also apply
tensile strains $\rm{\epsilon_{zig}}$ as mentioned in \sect{sec:sample_preparation} and
change the nature of the BZ. 
With the help of the harmonic lattice dynamics based GULP package, we obtain the
phonon dispersion curves along the symmetric $\Gamma$--M direction for samples with
tensile strain $\rm{\epsilon_{zig}} = 0.0, 0.9, 1.4, 3, 4, 7$ and 10\% as shown in \fig{fig05}.
We observe that the frequencies for the TA, LA, ZO, TO and LO modes decrease for a given
wavevector as we keep on increasing the strain; however, the ZA modes show the opposite trend. The flattening in frequencies occurs because the applied strain weakens the bond stiffness. 

\begin{figure}[htp]
    \centering
    \hspace{0.1cm}
    \subfloat[$\rm{{v}_{g}}$ ]{\includegraphics[width=0.44\textwidth]{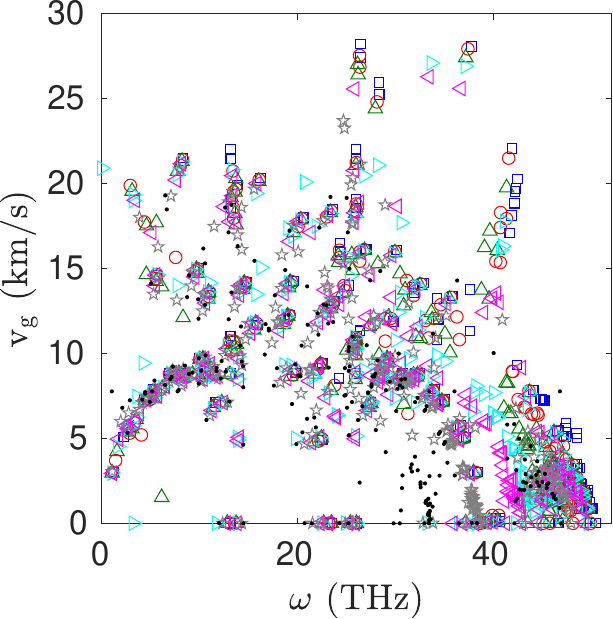}\label{fig06a}}
     \hspace{0.1cm}
    \subfloat[$\tau$ ]{\includegraphics[width=0.45\textwidth]{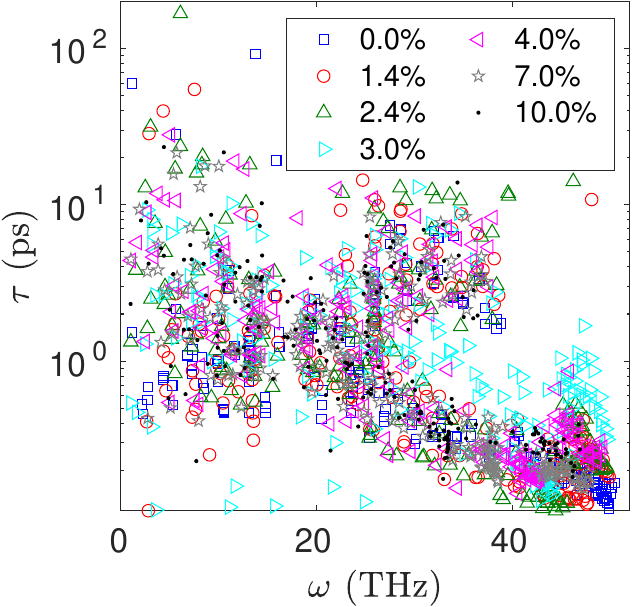}\label{fig06b}}       
    \caption{(a) $\rm{v_g}$ (km/s) and (b) $\tau$ (ps) comparison of unstrained and strained pristine graphene samples for all allowed phonon modes of the $1^{st}$ quadrant of the first BZ.}
  \label{fig06}
\end{figure}

We also plot group velocities and phonon lifetime for all allowed phonon modes of the $1^{st}$ quadrant of the first BZ in \figs{fig06a} and \figx{fig06b}, respectively, for the unstrained and strained samples. Following the symmetry of the BZ, they are representatives of the whole first BZ as well.
\begin{figure}[htp]
    \centering
    \hspace{0.1cm}
    \subfloat[$\rm{v_{{g}_{rms}}^2}$]{\includegraphics[width=0.45\textwidth]{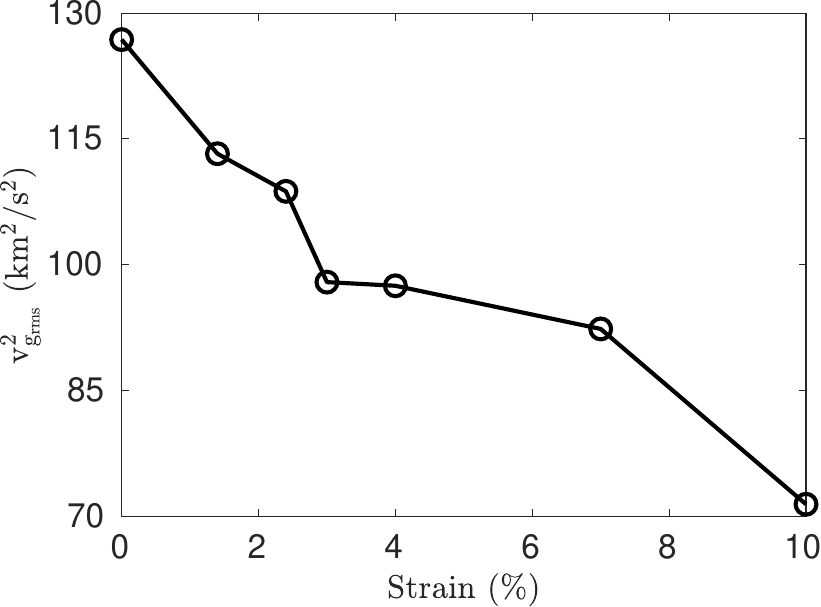}\label{fig07a}}
    \subfloat[$\rm{\overline{\tau}}$]{\includegraphics[width=0.45\textwidth]{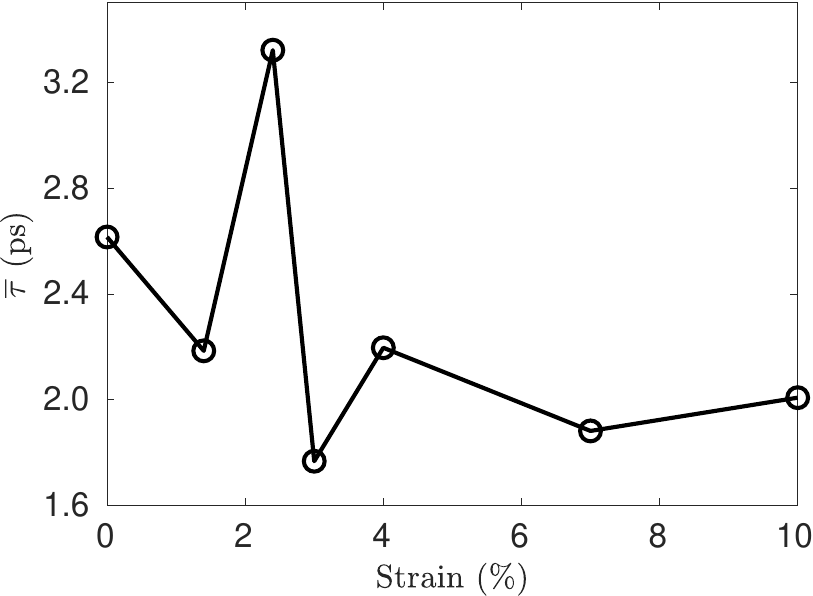}\label{fig07b}}\\    
    \subfloat[\rm{$\overline{l}$}]{\includegraphics[width=0.45\textwidth]{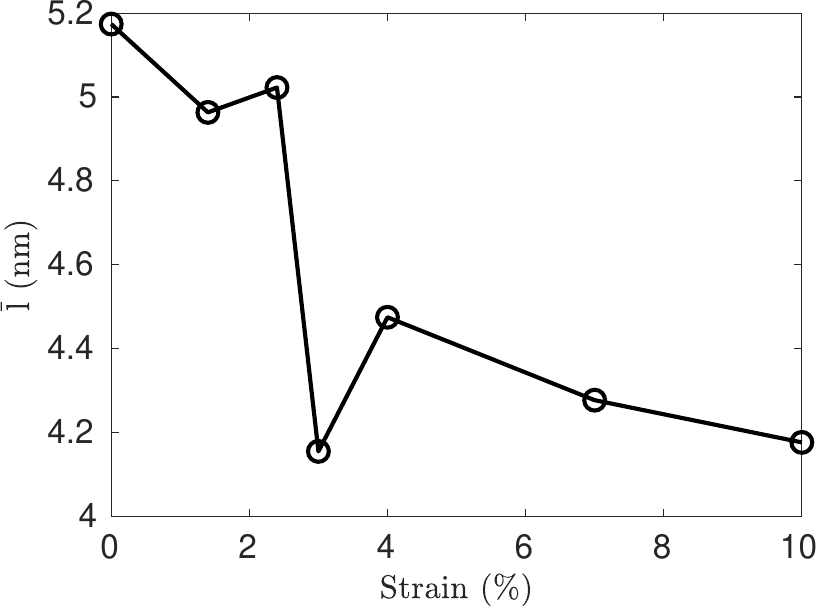}\label{fig07c}}
    \subfloat[k]{\includegraphics[width=0.45\textwidth]{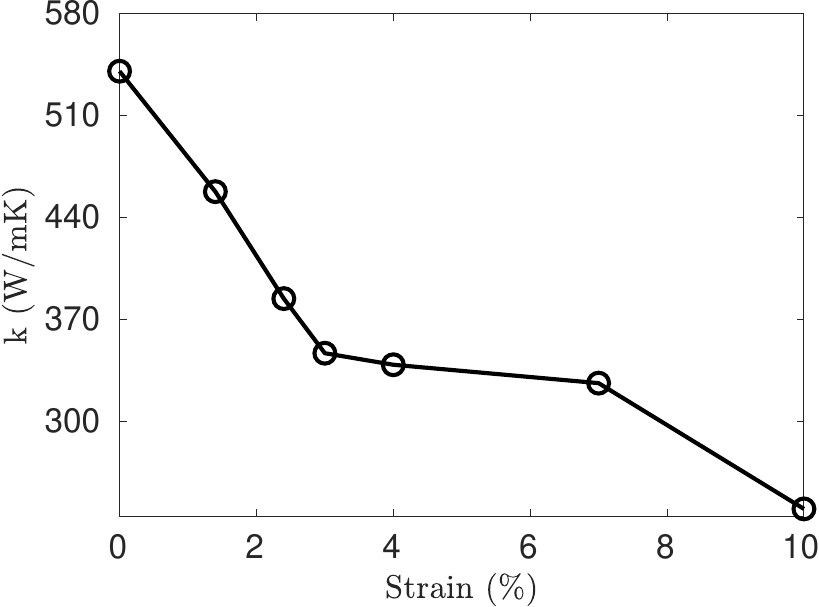}\label{fig07d}}
    \caption{ Considering the whole first BZ, (a) $\rm{v_{g_{rms}}^2}~ (km^2/s^2)$  (b) $\overline{\tau}$ (ps) (c) \rm{$\overline{l}$} (nm) and (d) $\kappa$ (W/mK) comparison of unstrained and strained pristine graphene samples of size $31 \times 26 $~\AA$^2$.}
  \label{fig07}
\end{figure}
Although the values overlap for different strains in these plots, there exist some remarkable trends. For strained graphene samples, some modes show null $\rm{v}_g$ values.
We plot the square of the root mean square of the group velocities,
$\rm{v_{{g}_{rms}}^2}$, in \fig{fig07a} and the mean of the phonon lifetimes, $\bar{\tau}$, in \fig{fig07b} for all allowed phonon modes in the whole of the first BZ to show some visible trends.
A monotonically decreasing trend can be seen for $\rm{v_{{g}_{rms}}^2}$ in \fig{fig07a}, thus confirming our earlier observation that the dispersion curves start to flatten as the strain increases. We observe almost 46\% decrease in  $\rm{v_{{g}_{rms}}^2}$ when the strain increases from 0 to 10\%.
However, the scattering processes that dictate the lifetime $\tau$ are not as sensitive to strain. The mean lifetime fluctuates between 0 and the 10\% strain as shown in \fig{fig07b}, but the overall magnitude appears to decline for higher strain values. Moreover, from \fig{fig06b}, we can also observe that the distribution of $\tau$ shows a decreasing trend for all strains as the frequency increases. Near the BZ center, the phonon lifetimes of the ZA modes are found to be as high as $100$ ps. This is because the phonon density of states (PDOS), as shown in \fig{fig08}, is very small near the center of the zone.
\begin{figure}[htb!]
\centering
\includegraphics[width=0.5\textwidth]{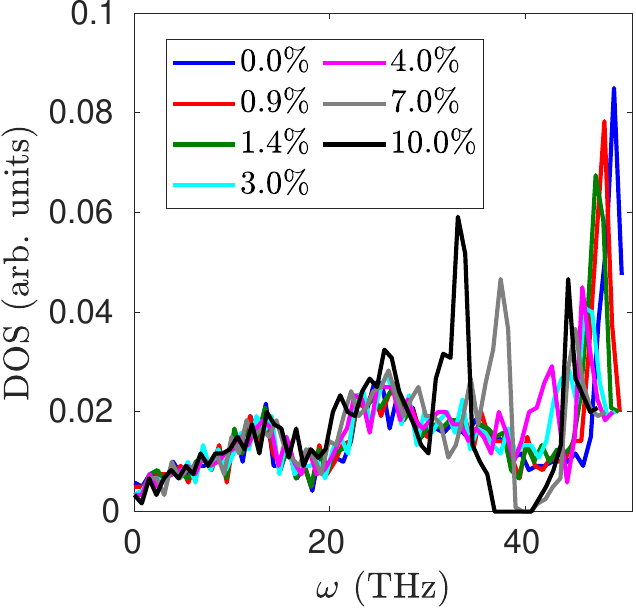}
  \caption{DOS comparison for different strains.}
  \label{fig08}
\end{figure}
This means that less modes are available for scattering to occur, which results in higher phonon lifetimes. Apart from BZ centers, the maximum distribution of $\tau$ was found between 20 and 40 THz frequency (\cite{qiu2012reduction}) for all strains because the PDOS is higher in this frequency range. In \fig{fig08}, it is also observed that the PDOS distribution shifts toward lower frequencies as the strain increases. The mean free path (mfp) is given by $\rm{\overline{l}} = \rm{v_{g_{rms}}} \times \bar{\tau}$, which shows an overall decreasing trend in \fig{fig07c} as the strain increases. This occurs because of the decreasing trend of $\rm{v}_g$ and the almost constant behavior of $\bar{\tau}$.

Subsequently, once the mode properties $\tau$ and $\rm{v}_g$ are obtained, we can calculate the total TC of the unstrained and strained graphene samples according to \eqn{kSED}, where the whole first BZ is considered. As shown in \fig{fig07d}, the TC continues to decrease as the strain increases~\cite{wei2011strain, park2017limited}.  The TC values are 540, 458, 384, 347, 339, 326 and 240 W/mK for 0.0\%, 0.9\%, 1.4\%, 3.0\%, 4.0\%, 7.0\% and 10.0\% strains, respectively. Between
0 and 3\% strain, we find that the TC decreases by approximately 36\%. For the same strain
difference, $\rm{v_{{g}_{rms}}^2}$ decreases by around 25\%.
Between strains 3\% and 10\%, TC decreases by approximately $31\%$, which is very close to the approximately 30\% decrease found for the sample $150 \times 100$~\AA$^2$ studied with the Green-Kubo method. 

\subsection{Quantifying the ripple effect on TC}\label{sec:results_ripple}
To distinguish between the effects of ripple and strain on thermal conductivity, we study two different sizes of pristine graphene samples under $0\%$ to $10\%$ tensile strain: (a) $150 \times 100~$\AA$^2$~ with the Green-Kubo method, and (b)
 $31 \times 26 $~\AA$^2$~ with the SED method. As both are equilibrium MD-based methods, the total percentage changes in TC should be close to each other. 
As mentioned above, between strains 3\% and 10\%, both samples with different methods show a similar decrease in TC, 31\% with SED and 30\% with Green-Kubo. In this strain range, there are no ripples for any of the two samples. 
Thus, it can be said that the strain has the same effect on the TC of the ripple-free systems.
However, the smaller sample without any ripples shows an approximately 36\% reduction in TC between 0 and 3\% strain. The strain effect should be the same on both samples and therefore the same percentage reduction should be observed for the $150 \times 100 $~\AA$^2$~  sample between 0-3\% strain. However, as mentioned earlier, the
combined effect of ripples and strain in this 0-3\% strain range shows an overall
increase of 25\% in the value of TC.  We can hypothesize that the effects of ripples and strain on the TC can be linearly decoupled, that is,
\begin{equation}\label{combined_effect}
    \left.{\frac{\Delta \kappa}{\kappa}} \right| _{overall}  = \left.{\frac{\Delta \kappa}{\kappa}} \right| _{strain}  +    \left.{\frac{\Delta \kappa}{\kappa}}\right|_{ripple}
\end{equation} 
Using \eqn{combined_effect}, we have $ \left.{\frac{\Delta \kappa}{\kappa}}\right|_{ripple} =  \left.{\frac{\Delta \kappa}{\kappa}} \right| _{overall} -
\left.{\frac{\Delta \kappa}{\kappa}} \right| _{strain} = +25\% - (-36\%) = +61\%$.
Therefore, we stipulate that ripples are responsible for approximately 61\% increase in TC for unstrained samples. This effectively means that the TC of the $150 \times 100 $~\AA$^2$~ sample without ripples  should be 1.61 times the TC of the same sample with ripples. This means that the ripple-free sample should have $\kappa = 785 \times 1.61 \approx 1264$~ W/mK. In other words, this analysis helps to show that ripples arising from thermal fluctuations at room temperature reduce the TC of the
$150 \times 100 $~\AA$^2$~ sample by approximately 61\%. The same analysis could be performed for other sample sizes at different temperatures to obtain a full picture of
the effect of the ripples on the TC. For lack of time, we leave this analysis to be done in our future work.
\section{Effect of curvature on thermal conductivity}\label{sec:curvature}
\begin{figure}[htb!]
\centering
\includegraphics[width=0.75\textwidth]{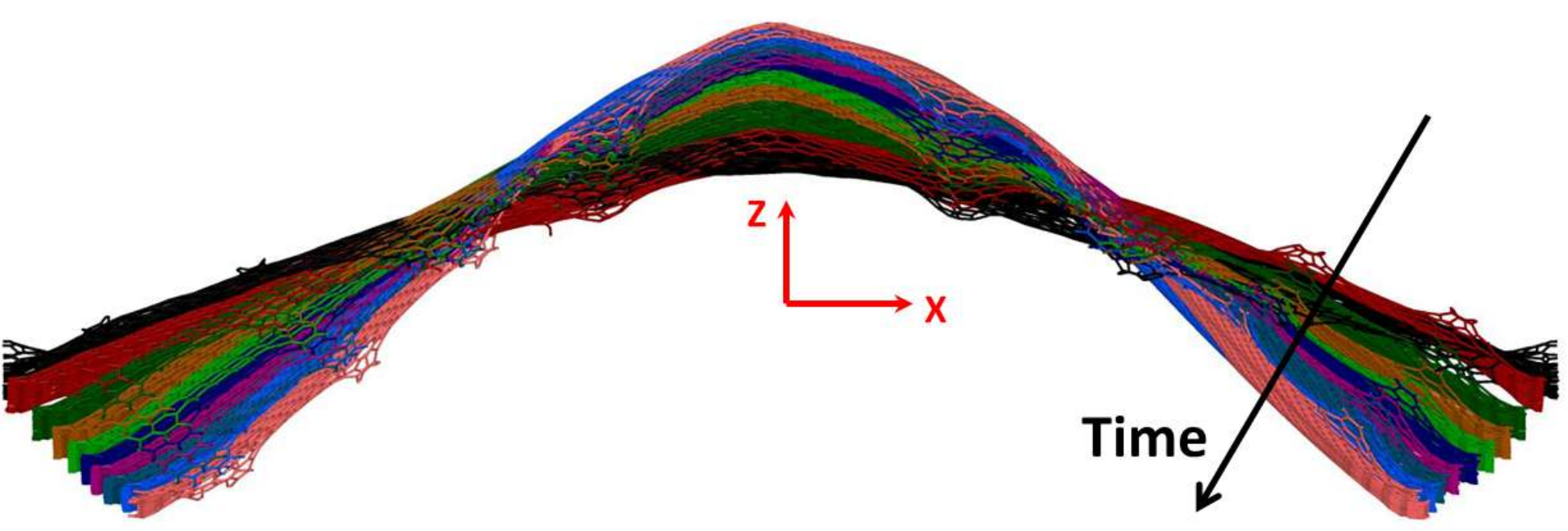}
  \caption{Front view of a graphene sheet with different curvatures}
  \label{fig09}
\end{figure}

The bending of the graphene sheet with grain boundaries could be another way through which the TC can be modulated, and we achieve this through the following procedure.
We prepare graphene samples of size $150 \times 100 $~\AA$^2$~ containing GBs as mentioned in \sect{sec:sample_preparation} and \fig{fig01}. Then we perform the simulation steps mentioned in \sect{sec:greenkubo}. However, during the NPT run, we use different pressure conditions. Instead of simulating at zero pressure, we allowed the atoms to evolve under a pressure of 200 bar applied along the x-axis. As the system evolves in time, graphene sheets start to bend and develop curvatures as shown in \fig{fig09}.
The figure presents the side view of the sheet over the NPT runtime of $0.9$~ns, where
9 different configurations are extracted at the interval of 100~ps. 
Note that pristine graphene samples without GBs do not develop curvatures during a similar NPT run. The strain field caused by the grain boundary is responsible for this curvature. Earlier studies~\cite{mortazavi_2012,li_2013,zhang_2017,zhuang_2019}
had developed the curvature in the pristine graphene sheet geometrically. 
Our technique of generating curvature in the graphene sheets with GBs, where the curvature is induced by the NPT ensemble under specific pressure conditions, is unique and has never been reported in the literature. 

We also calculate the curvature of curved graphene samples with the help of the
second fundamental form of a smooth surface on the tangent plane in the three-dimensional Euclidean space~\cite{banchoff:2022}. 
\begin{figure}[htb!]
    \centering
    \subfloat[$\theta = 18.73^\circ$, $\rho = 0.0167~\text{\AA}^{-1}$]{\includegraphics[width=0.47\textwidth]{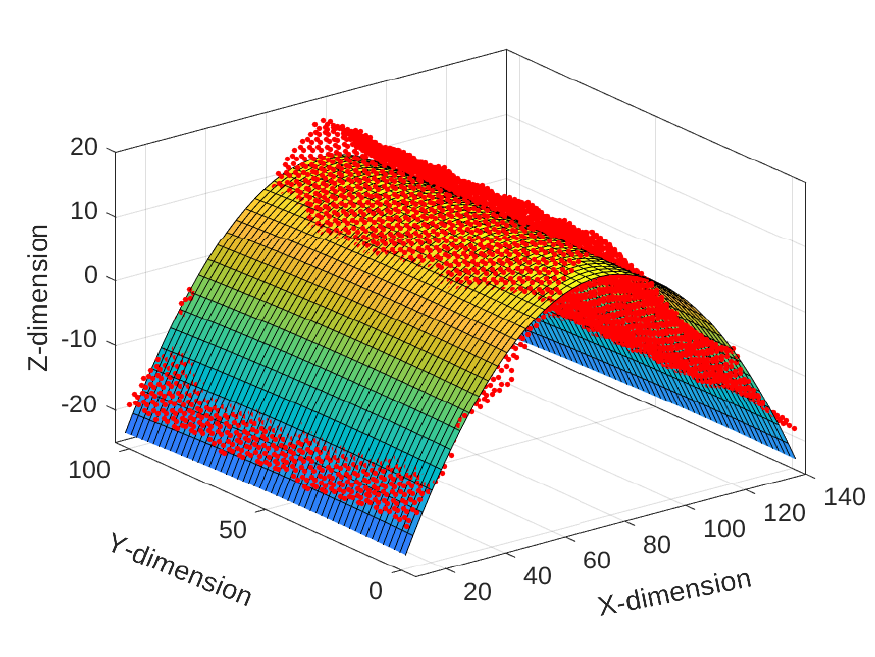}}\label{fig10a}
    \subfloat[$\theta = 32.2^\circ$, $\rho = 0.0162~\text{\AA}^{-1}$]{\includegraphics[width=0.47\textwidth]{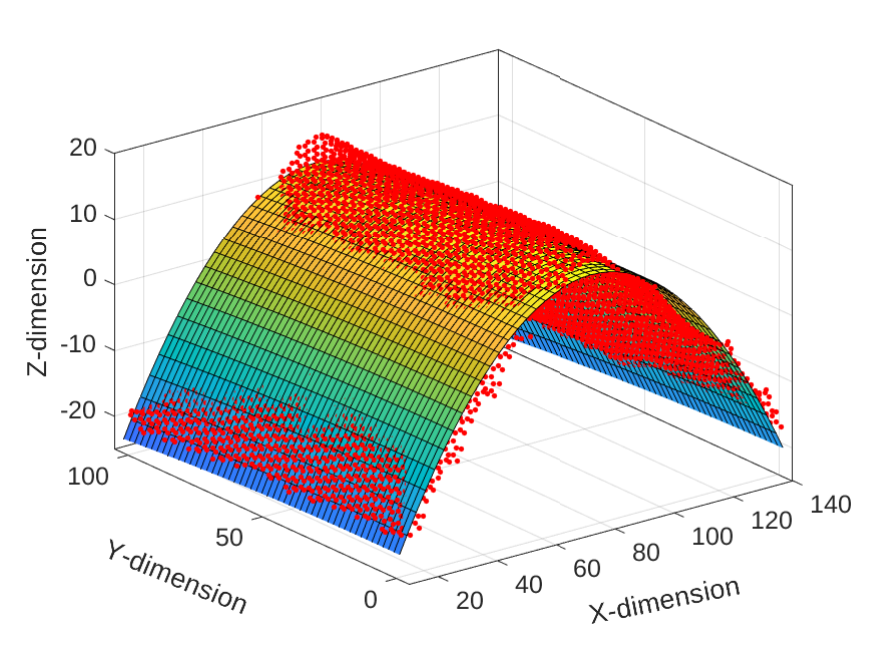}}\label{fig10b}
    \caption{Second order polynomial surface fitting to graphene surfaces with GB tilt angles $18.73^\circ$ and $32.2^\circ$.}
    \label{fig10}
\end{figure}
We consider two specific graphene samples of size $150 \times 100$~\AA$^2$~ with GB tilt angles $\theta = 18.73^\circ$ and $\theta = 32.2^\circ$ and regularize their
irregular surfaces shown in \fig{fig10} by fitting their z coordinates with second-degree polynomial surfaces represented by \eqn{surface_fit} as given below:
\begin{equation}
   z = f(x, y) = A + Bx + Cy + Dx^2 + Exy + Fy^2, \label{surface_fit}
\end{equation}
where $A, B, C, D, E$ and $F$ are polynomial coefficients. 
At the middle point of the smoothened surface, the normal
vector to the surface is along the z-axis, and the tangent plane is horizontal. 
In this case, the second fundamental form of the surface $z=f(x,y)$ is given in terms of Hessian $\mathcal{H}$ of $f$. This Hessian is given by
\begin{align}
\mathcal{H}= 
\begin{bmatrix}
  f_{xx} & f_{xy} \\
  f_{xy} & f_{yy}
\end{bmatrix}
=
\begin{bmatrix}
  2D & E \\
  E & 2F
\end{bmatrix}, \label{hessian}
\end{align}
 whose eigenvalues are the principal curvatures~\cite{banchoff:2022}, which represent the maximum and the minimum curvatures of the surface at the point where the tangent plane is horizontal. We calculate the maximum curvatures, $\rho \equiv \rm{max\, eigenvalue\, of\,}  \mathcal{H} $, for all
configurations shown in \fig{fig09} over an NPT runtime of 0.9~ns and find that they
continue to increase from $\rho = 0$ to $\rho =0.026$~\AA$^{-1}$. The minimum
curvatures remain close to zero for all configurations for any tilt angle. It is interesting to find that the principal curvatures do not depend on the GB tilt angles.

\begin{figure}[htb!]
    \centering
    \subfloat[$\theta = 18.73^\circ$]{\includegraphics[width=0.45\textwidth,height=2.2in]{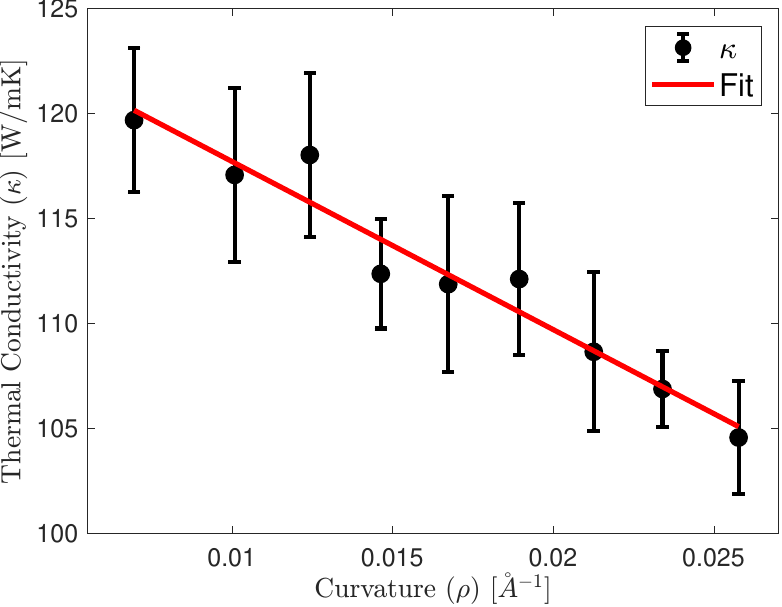}}\label{fig11a}
    \subfloat[$\theta = 32.2^\circ$]{\includegraphics[width=0.45\textwidth,height=2.2in]{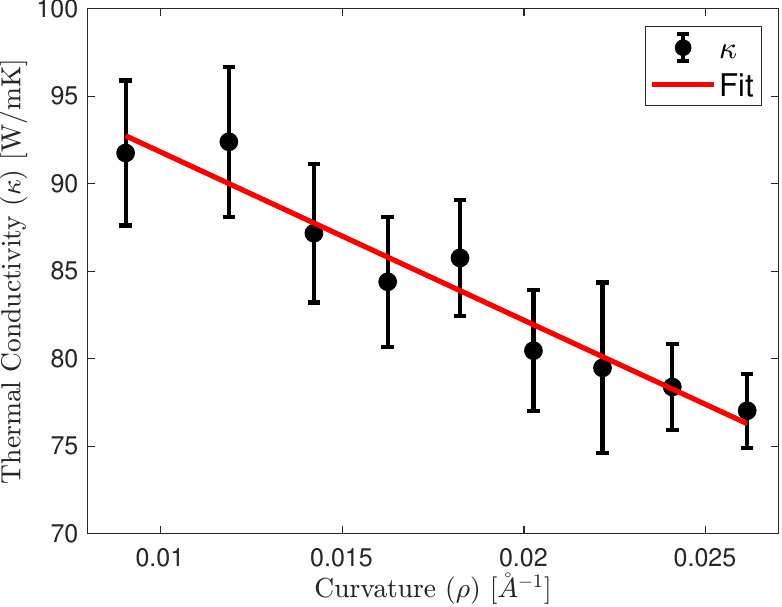}}\label{fig11b}
    \caption{Dependence of thermal conductivity on maximum curvature values for
    $150 \times 100 $~\AA$^2$ graphene with two GB tilt angles $\theta = 18.73^\circ$ and $\theta = 32.2^\circ$.}
    \label{fig11}
\end{figure}

Once we obtain the curved configurations as shown in \fig{fig09} and calculate the maximum curvatures $\rho$, we run the Green-Kubo simulations as mentioned in \sect{sec:greenkubo} to obtain their TC values. 
For graphene samples of size $150 \times 100$~\AA$^2$, regardless of their curvatures, we have taken the volume of
the system in \eqn{discreteGKeqb} as $V = 150 \times 100 \times 3.34$~\AA$^3$, where 3.34~\AA~ is the interlayer spacing. The dependence of the TC values ($\kappa$) on $\rho$ has been shown in \fig{fig11} for $150 \times 100 $~\AA$^2$ graphene with two GB tilt angles $\theta = 18.73^\circ$ and $\theta = 32.2^\circ$.
Here, $\kappa$ is the average of the in-plane diagonal components $\kappa_{\rm{xx}}$ and $\kappa_{\rm{yy}}$, and we find that $\kappa_{\rm{zz}}$ remains close to zero for all curvatures.
It can be seen from \fig{fig11} that the TC values decrease almost linearly with
increase in $\rho$. The error bar represents the standard error of the TC values of five replicas belonging to the same NVT ensemble (see \sect{sec:greenkubo} for more details of the standard error). We fit the data with straight lines and obtain following relationships between $\kappa$ and $\rho$ for the two tilt angles:
\begin{align}
\begin{split}
    \kappa &= 125.7- 800.3\,\rho, \quad \textrm{for} \quad \theta=18.73^\circ,\\
     \kappa &= 101.4- 961.9\,\rho, \quad \textrm{for} \quad \theta=32.2^\circ.
    \label{curvature_fit} 
    \end{split}
\end{align}
This clearly shows that the rate of decrease in TC with curvature depends on the tilt angle, which is greater for $\theta = 32.2^\circ$ than for $\theta = 18.73^\circ$. Between $\rho = 0.007$~\AA$^{-1}$ and $\rho = 0.026$~\AA$^{-1}$, we also find that the mean TC values decreased by approximately 13\% for $\theta = 18.73^\circ$ when the TC decreased from 120~W/mK to 105~W/mK, and by approximately 16\% for $\theta = 32.2^\circ$ when the TC decreased from 92~W/mK to 77~W/mK. 
The percentage reduction is close to the values reported by Xiangjun Liu et al.~\cite{liu_2019} for silicon nanowires; however, these reductions in TC values due to curvature differ from other works~\cite{mortazavi_2012,li_2013, zhang_2017, zhuang_2019} who introduced different kinds of geometrically prepared curvature in different samples.

\begin{figure}[htp]
    \centering
  \includegraphics[width=0.55\textwidth,height=2.2in]{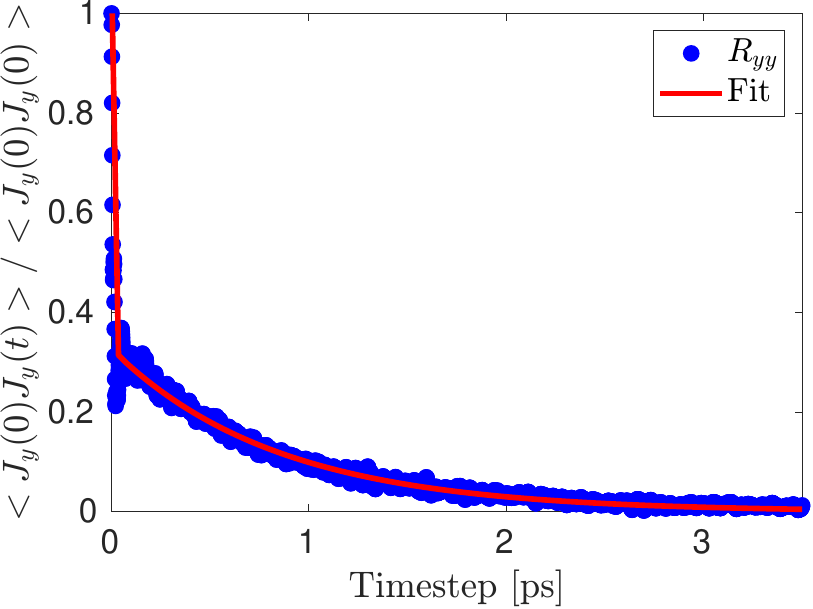}
    \caption{A typical representation of a fitted curve to the normalized HCACF $R_{\rm{yy}}$ for $\theta = 18.73^\circ$ at $\rho = 0.019\,\text{\AA}^{-1}$.}
    \label{fig12}
\end{figure}

For graphene with grain boundaries that exhibit curvatures, we also investigated the reasons behind the TC trends with respect to curvature, as shown in \fig{fig11}. For this purpose, we focus on the ensemble average of the heat current autocorrelation functions (HCACF), $\avg{J_{x}(t)J_{x}(0)}$ and $\avg{J_{y}(t)J_{y}(0)}$. 
They almost decay exponentially in accordance with the macroscopic relaxation law and Onsager's postulate for microscopic thermal fluctuations~\cite{che_2000}, and can be fitted with
double exponential decay functions~\cite{che_2000} in the following way: 
\begin{align}
\begin{split}
   R_{\rm{xx}}=\frac{\avg{J_x(t)J_x(0)}}{\avg{J_x(0)J_x(0)}} = A_{1_x} e^{\frac{-t}{\tau_{1_x}}} + A_{2_x} e^{  \frac{-t}{\tau_{2_x}}}, \quad 
   R_{\rm{yy}}=\frac{\avg{J_y(t)J_y(0)}}{\avg{J_y(0)J_y(0)}} = A_{1_y} e^{\frac{-t}{\tau_{1_y}}} + A_{2_y} e^{  \frac{-t}{\tau_{2_y}}},
   \end{split}
   \label{exponential_hcacf}
\end{align}
where $R_{\rm{xx}}$ and $R_{\rm{yy}}$ are the normalized HCACFs. The dimensionless constants $A_{1_{x,y}}$ and $A_{2_{x,y}}$ are the fitting parameters, where the subscripts
$x$ and $y$ are used for the normalized HCACFs along the x and y axes. The parameters $\tau_{1_{x,y}}$ and $\tau_{2_{x,y}}$ 
can be interpreted as measures of phonon scattering or relaxation times. 
For illustration, we show the fitting of $R_{\rm{yy}}$ for $\theta = 18.73^\circ$ at $\rho = 0.019$~\AA$^{-1}$ for a replica of the NVT ensemble in \fig{fig12},
 which gives $A_{1_y} = 0.75$, $A_{2_y}=0.33$, $\tau_{1_y}=0.01$~ ps and $\tau_{2_y}=0.84$~ ps. We present the estimated values of these parameters for different curvatures in Tables~\ref{table01} and ~\ref{table02} for graphene samples with GB tilt angles $\theta = 18.73^\circ$ and $\theta = 32.2^\circ$, respectively, for a replica of the NVT ensemble.
\begin{table}[htp]
\caption{\label{table01}Estimated parameters in \eqn{exponential_hcacf} for different curvatures for the graphene sample with the GB tilt angle $\theta = 18.73^\circ$ for a replica of the NVT ensemble.}
\centering
\begin{tabular}{ccccccccc}
\hline
$\rho$ [\AA$^{-1}$] & $A_{1_x}$ & $A_{2_x}$ & $\tau_{1_x}$ [ps] & $\tau_{2_x}$ [ps] & $A_{1_y}$ & $A_{2_y}$ & $\tau_{1_y}$ [ps] & $\tau_{2_y}$ [ps] \\
\hline
0.007 & 0.18 & 0.30 & 0.01 & 0.44 & 0.75 & 0.33 & 0.01 & 0.81 \\
0.010 & 0.26 & 0.30 & 0.01 & 0.43 & 0.75 & 0.33 & 0.01 & 0.83 \\
0.012 & 0.20 & 0.29 & 0.01 & 0.43 & 0.75 & 0.33 & 0.01 & 0.83 \\
0.015 & 0.15 & 0.29 & 0.01 & 0.43 & 0.75 & 0.33 & 0.01 & 0.82 \\
0.017 & 0.29 & 0.002 & 0.44 & 0.44 & 0.75 & 0.33 & 0.01 & 0.81 \\
0.019 & 0.25 & 0.26 & 0.01 & 0.51 & 0.75 & 0.33 & 0.01 & 0.84 \\
0.021 & 0.17 & 0.11 & 0.48 & 0.41 & 0.75 & 0.33 & 0.01 & 0.86 \\
0.023 & 0.81 & 0.23 & 0.01 & 0.56 & 0.75 & 0.33 & 0.01 & 0.83 \\
0.026 & 0.24 & 0.26 & 0.01 & 0.46 & 0.75 & 0.33 & 0.01 & 0.75 \\
\hline
\end{tabular}
\end{table}
\begin{table}[htp]
\caption{\label{table02}Estimated parameters in \eqn{exponential_hcacf} for different curvatures for the graphene sample with the GB tilt angle $\theta = 32.2^\circ$ for a replica of the NVT ensemble.}
\centering
\begin{tabular}{ccccccccc}
\hline
$\rho$ [\AA$^{-1}$] & $A_{1_x}$ & $A_{2_x}$ & $\tau_{1_x}$ [ps] & $\tau_{2_x}$ [ps] & $A_{1_y}$ & $A_{2_y}$ & $\tau_{1_y}$ [ps] & $\tau_{2_y}$ [ps] \\
\hline
0.009 & 0.04 & 0.31 & 0.01 & 0.40 & 0.78 & 0.32 & 0.01 & 0.57 \\
0.012 & 0.06 & 0.29 & 0.03 & 0.41 & 0.77 & 0.32 & 0.01 & 0.56 \\
0.014 & 0.09 & 0.29 & 0.01 & 0.42 & 0.77 & 0.32 & 0.01 & 0.56 \\
0.016 & 0.23 & 0.28 & 0.01 & 0.43 & 0.25 & 0.35 & 0.02 & 0.50 \\
0.018 & 0.33 & 0.26 & 0.01 & 0.44 & 0.29 & 0.34 & 0.01 & 0.52 \\
0.020 & 0.91 & 0.23 & 0.01 & 0.52 & 0.19 & 0.33 & 0.01 & 0.50 \\
0.022 & 0.16 & 0.15 & 0.17 & 0.70 & 0.77 & 0.33 & 0.01 & 0.52 \\
0.024 & 0.59 & 0.23 & 0.01 & 0.56 & 0.78 & 0.31 & 0.01 & 0.56 \\
0.026 & 0.10 & 0.26 & 0.01 & 0.41 & 0.50 & 0.33 & 0.01 & 0.50 \\
\hline
\end{tabular}
\end{table}
For $\theta = 18.73^\circ$, it can be seen from Table~\ref{table01} that $\tau_{1_x}$ is 0.01 ps or approximately 0.5 ps for different curvatures, $\tau_{2_x}$ varies narrowly between 0.43 ps and 0.56 ps, $\tau_{1_y}$ is always 0.01 ps and $\tau_{2_y}$ is around 0.8 ps. Remarkably, $A_{1_y}$ and $A_{2_y}$ do not change with curvature, whereas we do not find any
trends in $A_{1_x}$ and $A_{2_x}$. Similarly, for $\theta = 32.2^\circ$, it can be seen from Table~\ref{table02} that $\tau_{1_x}$ is 0.01, 0.03 or 0.17 ps for different curvatures, $\tau_{2_x}$ varies between 0.40 ps and 0.70 ps, $\tau_{1_y}$ is always 0.01 ps or 0.02 ps and $\tau_{2_y}$ is close to 0.5 ps or 0.6 ps. However, here only $A_{2_y}$ does not change much.

\begin{figure}[htp]
    \centering
    \subfloat[$\kappa_{xx,yy}^{\rm{derived}}$]{\includegraphics[width=0.45\textwidth]{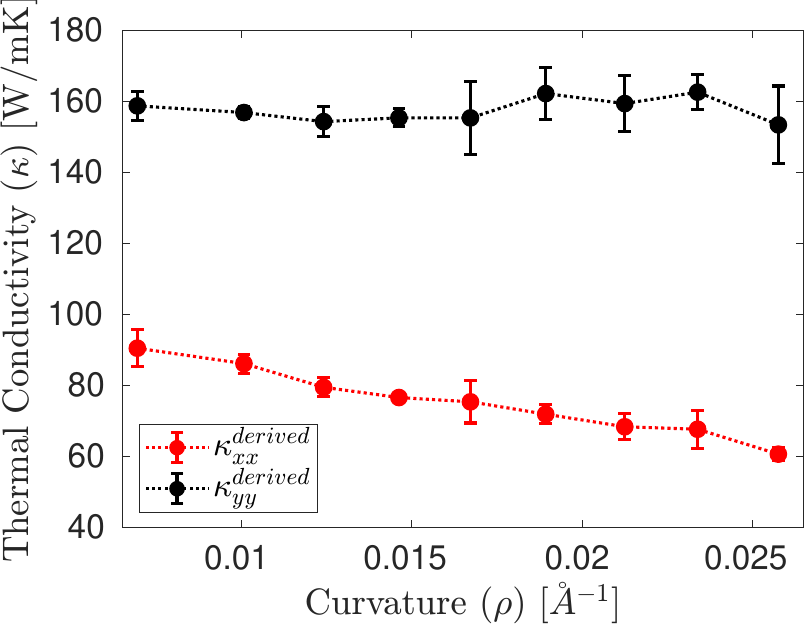}\label{fig13a}}\\
    \subfloat[$A_{1_x}\tau_{1_{x,y}}+A_{2_{x,y}}\tau_{2_{x,y}}$] {\includegraphics[width=0.4\textwidth]{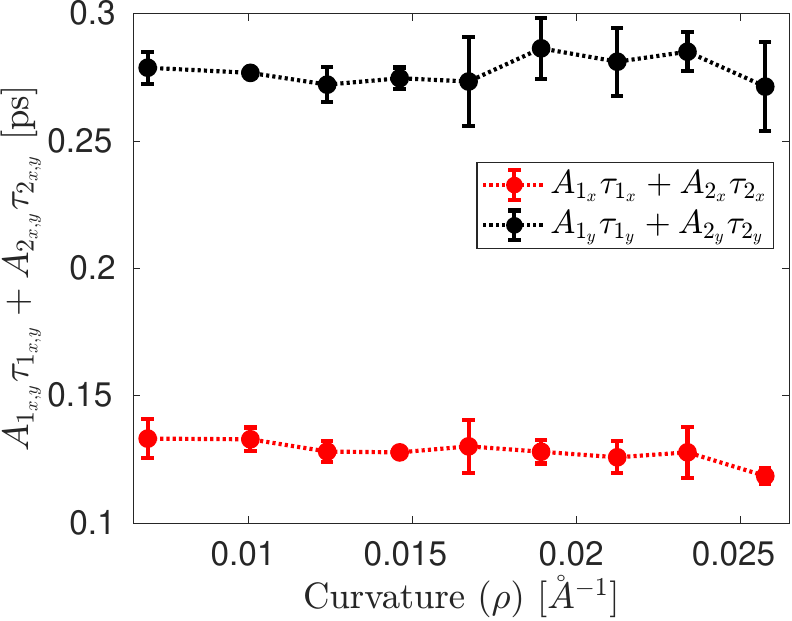}\label{fig13b}}
    \subfloat[$\avg{J_{x,y}(0)J_{x,y}(0)}$]{\includegraphics[width=0.4\textwidth]{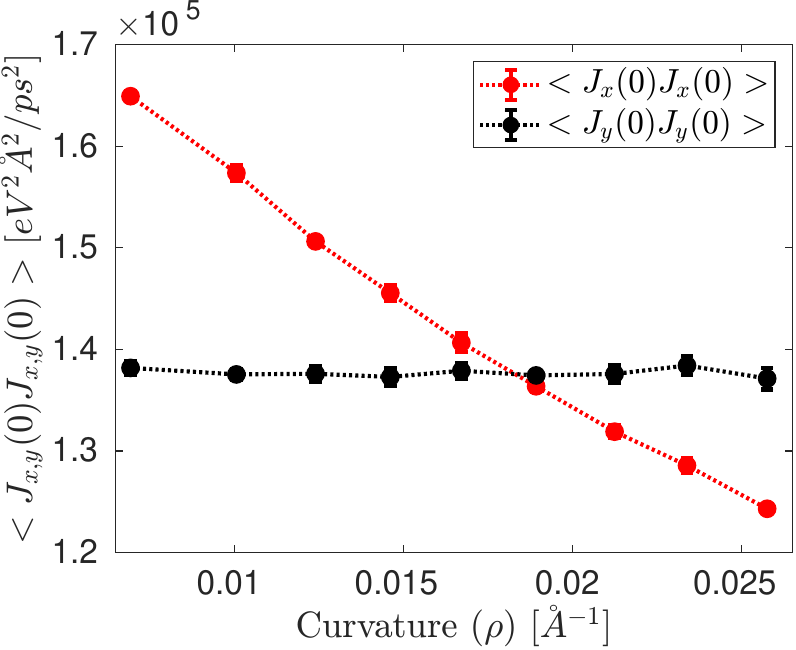}\label{fig13c}}
    \caption{For tilt angle $\theta = 18.73^\circ$, values of (a) 
    $\kappa_{\rm{xx}}^{\rm{derived}}$ and $\kappa_{\rm{yy}}^{\rm{derived}}$ in W/mK,
    (b) $A_{1_x}\tau_{1_x}+A_{2_x}\tau_{2_x}$ and $A_{1_y}\tau_{1_y}+A_{2_y}\tau_{2_y}$ in ps, and
    (c) $\avg{J_x(0)J_x(0)}$ and $\avg{J_y(0)J_y(0)}$ in eV$^2$\AA$^2$/ps$^2$
     for different curvatures. The standard errors in plot (c) are too small to be visible. The red color represents the x-component values and the black color
     represents the y-component values.}
    \label{fig13}
\end{figure}

\begin{figure}[htp]
    \centering
    \subfloat[$\kappa_{xx,yy}^{\rm{derived}}$]{\includegraphics[width=0.45\textwidth]{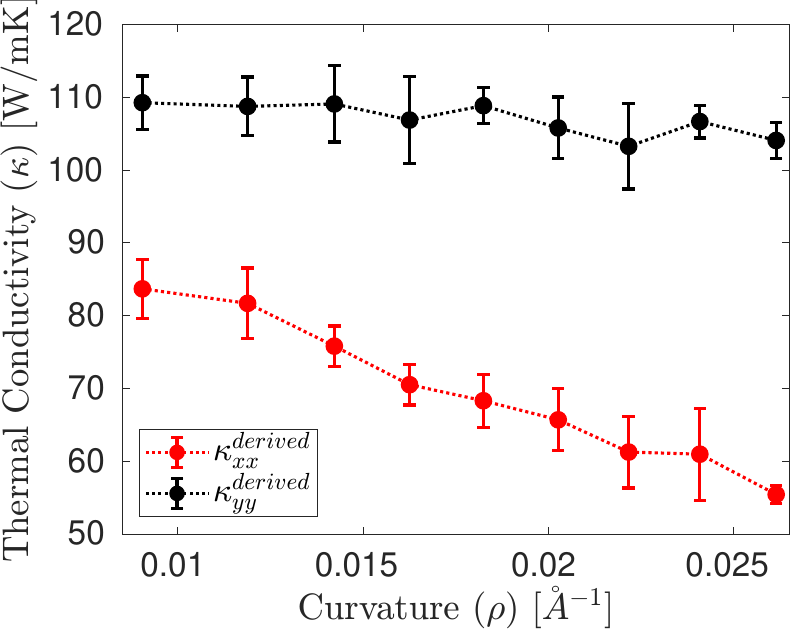}\label{fig14a}}\\
    \subfloat[$A_{1_x}\tau_{1_{x,y}}+A_{2_{x,y}}\tau_{2_{x,y}}$]{\includegraphics[width=0.4\textwidth]{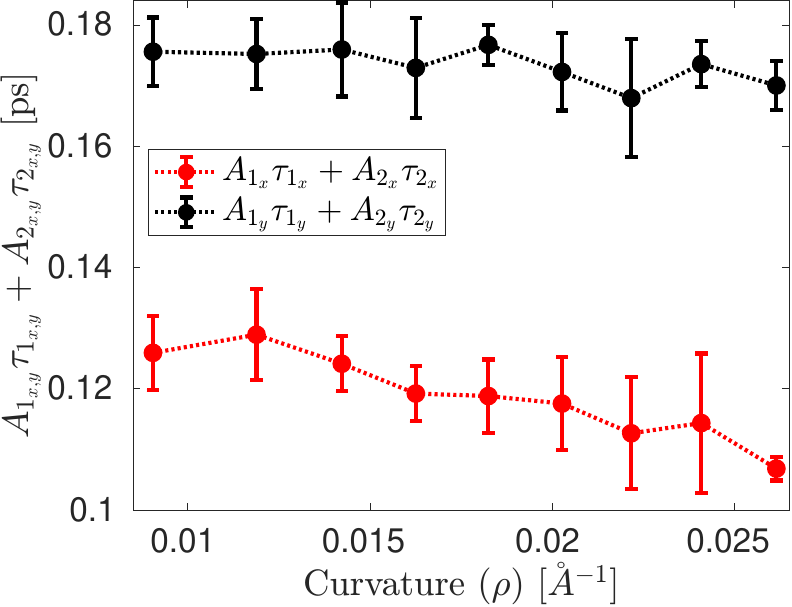}\label{fig14b}}
    \subfloat[$\avg{J_{x,y}(0)J_{x,y}(0)}$]{\includegraphics[width=0.4\textwidth]{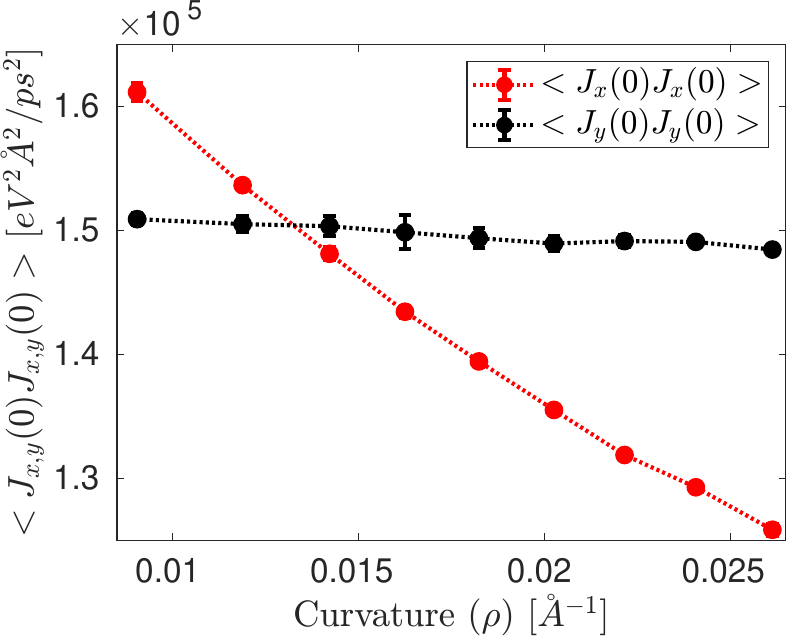}\label{fig14c}}
    \caption{For tilt angle $\theta = 32.2^\circ$, values of (a) 
    $\kappa_{\rm{xx}}^{\rm{derived}}$ and $\kappa_{\rm{yy}}^{\rm{derived}}$ in W/mK,
    (b) $A_{1_x}\tau_{1_x}+A_{2_x}\tau_{2_x}$ and $A_{1_y}\tau_{1_y}+A_{2_y}\tau_{2_y}$ in ps, and
    (c) $\avg{J_x(0)J_x(0)}$ and $\avg{J_y(0)J_y(0)}$ in eV$^2$\AA$^2$/ps$^2$
     for different curvatures. The standard errors in plot (c) are too small to be visible. The red color represents the x-component values and the black color
     represents the y-component values.}
    \label{fig14}
\end{figure}

\begin{figure}[htp]
    \centering
    \subfloat[$\theta = 18.73^\circ$]{\includegraphics[width=0.45\textwidth]{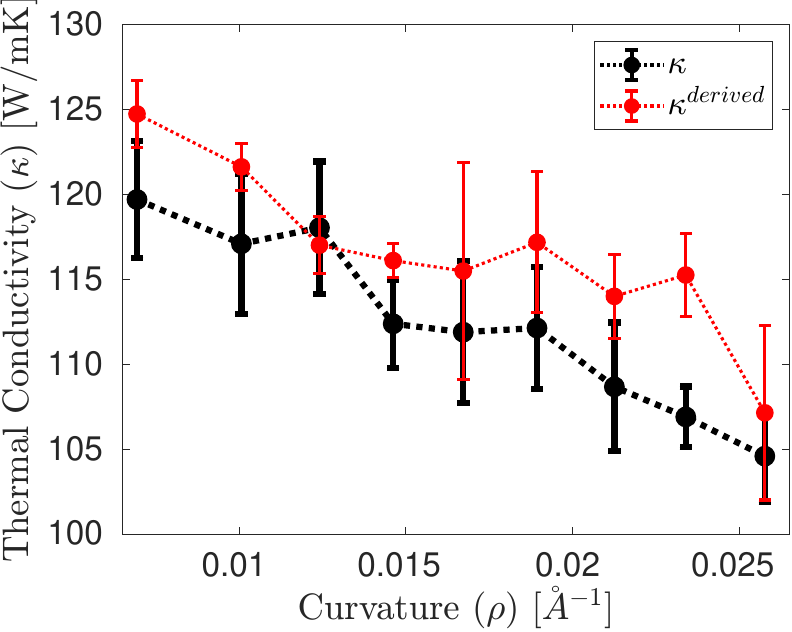}\label{fig15a}}
    \subfloat[$\theta = 32.2^\circ$]{\includegraphics[width=0.45\textwidth]{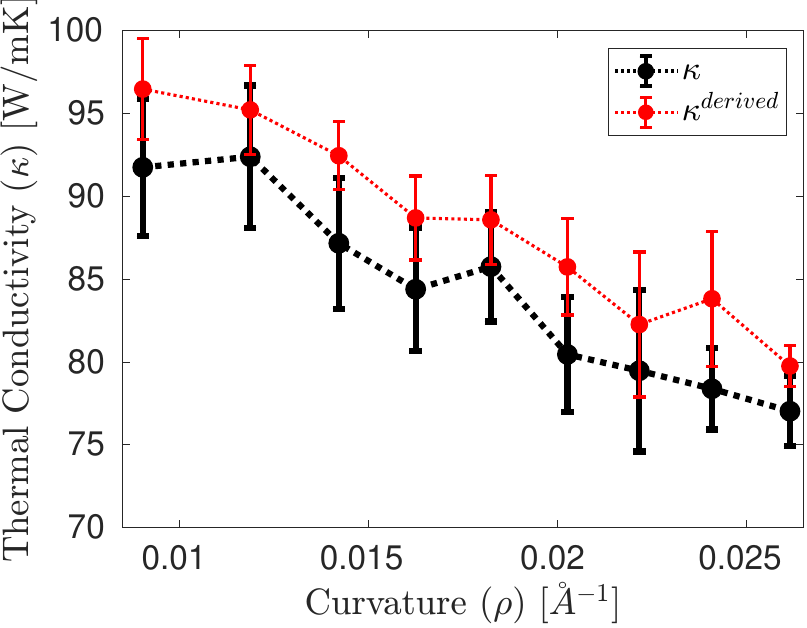}\label{fig15b}}
    \caption{Plots of $\kappa$ and  $\kappa^{\rm{derived}}$ vs  $\rho$ for GB tilt angles (a) $\theta = 18.73^\circ$ and (b) $\theta = 32.2^\circ$.}
    \label{fig15}
\end{figure}

In order to look for the underlying behavior of the TC with curvature, we plug \eqn{exponential_hcacf} into \eqn{GKeqb},  and obtain different measures for the TC
components $\kappa_{\rm{xx}}^{\rm{derived}}$ and $\kappa_{\rm{yy}}^{\rm{derived}}$ given by
\begin{align}
    \kappa_{\rm{xx}}^{\rm{derived}} = \frac{1}{V k_{\rm B} T^2}
    \left[\avg{J_x(0)J_x(0)}\left(A_{1_x}\tau_{1_x}+A_{2_x}\tau_{2_x}\right)\right], \quad 
    \kappa_{\rm{yy}}^{\rm{derived}} = \frac{1}{V k_{\rm B} T^2}
    \left[\avg{J_y(0)J_y(0)}\left(A_{1_y}\tau_{1_y}+A_{2_y}\tau_{2_y}\right)\right].
    \label{TC_components_derived}
\end{align}
The corresponding TC, $\kappa^{\rm{derived}}$, can be taken as the average of the two 
in-plane diagonal components, that is,  $\kappa^{\rm{derived}} = \frac{\kappa_{\rm{xx}}^{\rm{derived}}+\kappa_{\rm{yy}}^{\rm{derived}}}{2}$.
We plot $\kappa_{\rm{xx}}^{\rm{derived}}$ and $\kappa_{\rm{yy}}^{\rm{derived}}$ in \fig{fig13a} for the tilt angle $\theta = 18.73^\circ$ and find that 
$\kappa_{\rm{xx}}^{\rm{derived}}$  shows almost linearly decreasing trend; however, $\kappa_{\rm{yy}}^{\rm{derived}}$ remains constant with respect to the curvature. 
These TC components are mainly dependent on $A_{1_{c}}\tau_{1_{c}}+A_{2_{c}}\tau_{2_{c}}$ and $\avg{J_c(0)J_c(0)}$, where $c=\rm{x}$ or $\rm{y}$, whose plots with respect to curvature are presented in \figs{fig13b} and \figx{fig13c}, respectively.
We note that $A_{1_{c}}\tau_{1_{c}}+A_{2_{c}}\tau_{2_{c}}$ do not change much with curvature for both $c=\rm{x}$ and $\rm{y}$. 
Hence, the trends of $\avg{J_x(0)J_x(0)}$ and $\avg{J_y(0)J_y(0)}$ in \fig{fig13c} are strongly reflected in the trends of the TC components in \fig{fig13a}.
A similar exercise is followed for the tilt angle $\theta=32.2^\circ$, and here also,
we find the same linearly decreasing trend for $\kappa_{\rm{xx}}^{\rm{derived}}$ plotted in \fig{fig14a}.
However, $\kappa_{\rm{yy}}^{\rm{derived}}$ for $\theta=32.2^\circ$ plotted in \fig{fig14a} does not remain constant with curvature, but decreases almost linearly. For both x and y components, the linearly decreasing trends of $\avg{J_c(0)J_c(0)}$ and $A_{1_{c}}\tau_{1_{c}}+A_{2_{c}}\tau_{2_{c}}$ in \fig{fig14c} explain the decreasing trend of $\kappa_{\rm{cc}}^{\rm{derived}}$ in \fig{fig14a}, where $c=\rm{x}$ or $\rm{y}$.
There also exists a strong anisotropy in the TC components as the values of $\kappa_{\rm{xx}}^{\rm{derived}}$ are more than twice the values of $\kappa_{\rm{yy}}^{\rm{derived}}$ for different curvatures. 
In sum, only $\avg{J_c(0)J_c(0)}$ matters for explaining the trends in the TC component values for the GB tilt angle $\theta=18.73^\circ$, however, for $\theta=32.2^\circ$, both $\avg{J_c(0)J_c(0)}$ and $A_{1_{c}}\tau_{1_{c}}+A_{2_{c}}\tau_{2_{c}}$ matter.
Finally, we plot $\kappa^{\rm{derived}}$ for the two tilt angles in \fig{fig15}.
 It can be seen that the values of $\kappa^{\rm{derived}}$ are almost in agreement with the values of $\kappa$ for both tilt angles.
\section{Effect of different grain boundaries on thermal conductivity}\label{sec:GB}
\begin{figure}[H]
    \centering
  \includegraphics[width=0.5\textwidth]{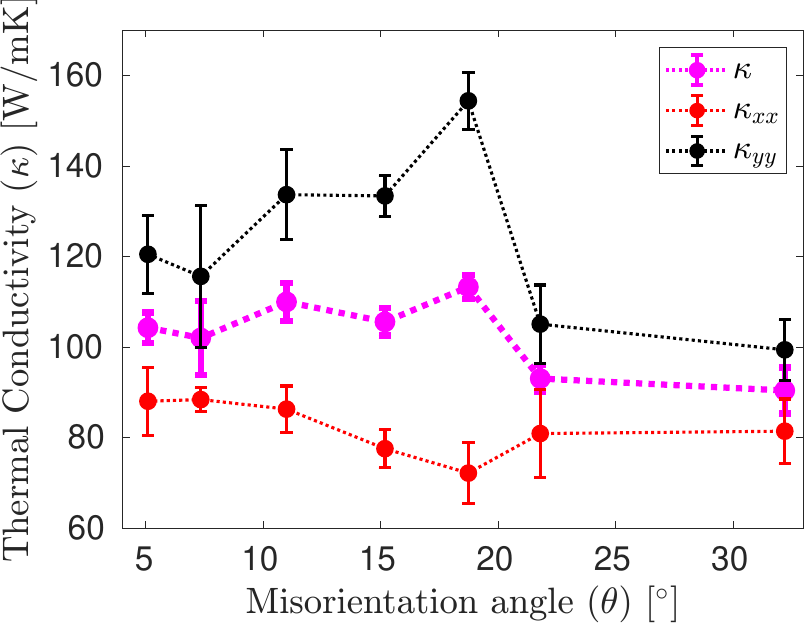}
    \caption{$\kappa_{\rm{xx}}$, $\kappa_{\rm{xx}}$, and $\kappa$ for all tilt angles ($\theta$ in degree)}
     \label{fig16}
\end{figure}

We have also studied the effect of different grain boundaries characterized by the tilt angles on the thermal conductivity. 
Samples with seven different tilt angles of approximately the same size were prepared according to the methodology described in \sect{sec:sample_preparation}. The TCs of the samples were calculated with the Green-Kubo method as described in \sect{sec:greenkubo} and the values are plotted in \fig{fig16}.
Similarly to Abhikeern and Singh~\cite{abhikeern2024latticethermalconductivityphonon}, 
our TCs strongly depend on the tilt angles, and there exists strong anisotropy along the x and y axes in the TCs. 

We also fitted the normalized HCACFs, $R_{\rm{xx}}$ and $R_{\rm{yy}}$, with the double exponential functions in \eqn{exponential_hcacf} and estimated the parameters values for different tilt angles in Tables~\ref{table03} for a replica of the NVT ensemble.
\begin{table}[htp]
\caption{\label{table03}Estimated parameters in \eqn{exponential_hcacf} for different tilt angles for a particular replica of the NVT ensemble.}
\centering
\begin{tabular}{ccccccccc}
\hline
$\theta [{}^\circ]$ & $A_{1_x}$ & $A_{2_x}$ & $\tau_{1_x}$ [ps]& $\tau_{2_x}$ [ps]& $A_{1_y}$ & $A_{2_y}$ & $\tau_{1_y}$ [ps]& $\tau_{2_y}$ [ps] \\
\hline
5.09 & 0.22 & 0.34 & 0.03 & 0.49 & 0.77 & 0.32 & 0.01 & 0.65 \\
7.34 & 0.78 & 0.27 & 0.01 & 0.61 & 0.82 & 0.30 & 0.01 & 0.66 \\
10.99 & 0.35 & 0.27 & 0.01 & 0.47 & 0.84 & 0.27 & 0.01 & 0.77 \\
15.18 & 0.89 & 0.25 & 0.01 & 0.51 & 0.81 & 0.28 & 0.01 & 0.78 \\
18.73 & 0.46 & 0.24 & 0.01 & 0.56 & 0.77 & 0.30 & 0.01 & 0.84 \\
21.79 & 0.25 & 0.22 & 0.01 & 0.45 & 0.32 & 0.25 & 0.01 & 0.5 \\
32.20 & 0.87 & 0.26 & 0.01 & 0.49 & 0.77 & 0.32 & 0.01 & 0.55 \\
\hline
\end{tabular}
\end{table}
As argued above in \sect{sec:curvature}, we note that $\tau_{1_x}$ and $\tau_{1_y}$ are 0.01 ps for almost all tilt angles, $\tau_{2_x}$ varies narrowly between 0.45 ps and 0.61 ps, and $\tau_{2_y}$ varies between 0.5 ps and 0.84 ps. We do not find any
trends in $A_{1_{x,y}}$ and $A_{2_{x,y}}$.
\begin{figure}[H]
    \centering
    \subfloat[$\kappa^{derived}$]{\includegraphics[width=0.47\textwidth]{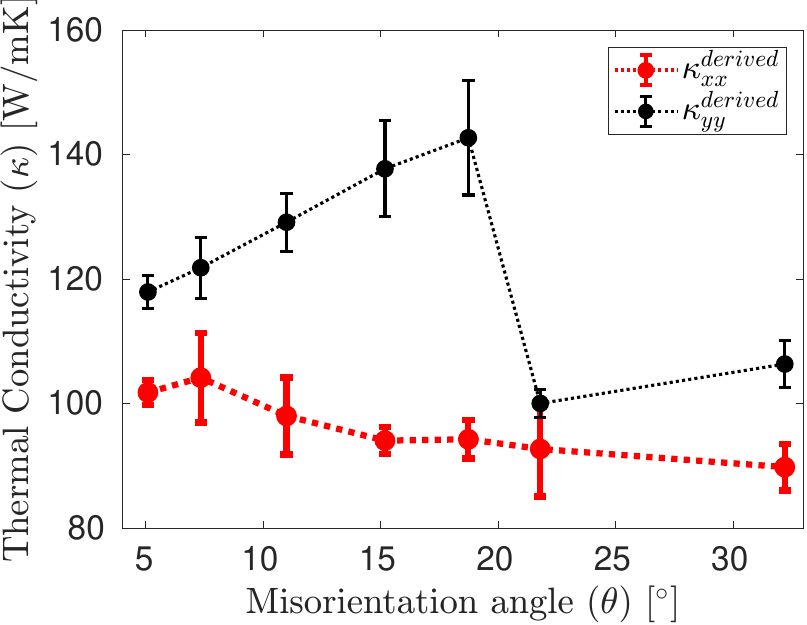} \label{fig17a}}
    \subfloat[$A_{1x,y}\tau_{1x,y} + A_{2x,y}\tau_{2x,y}$]{\includegraphics[width=0.45\textwidth]{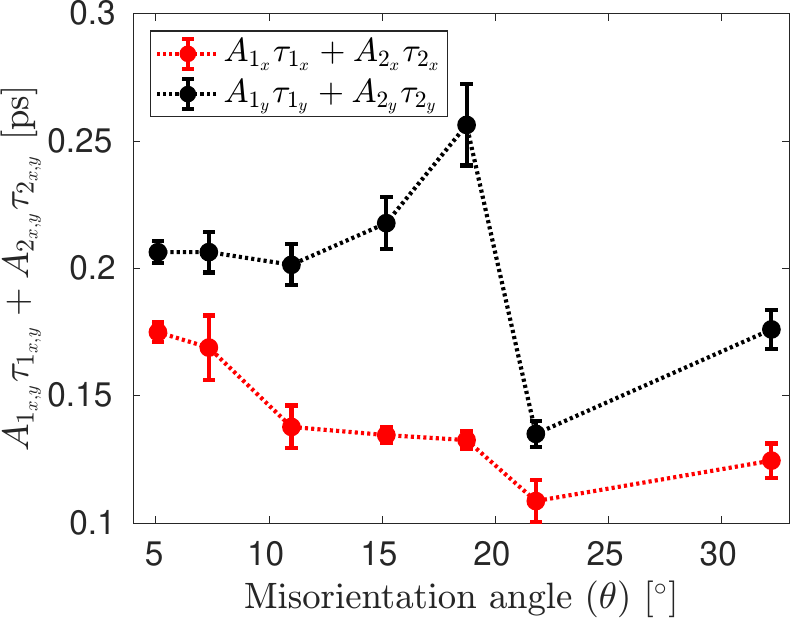} \label{fig17b}}\\
    \subfloat[$\avg{J_{x,y}(0)J_{x,y}(0)}$]{\includegraphics[width=0.45\textwidth]{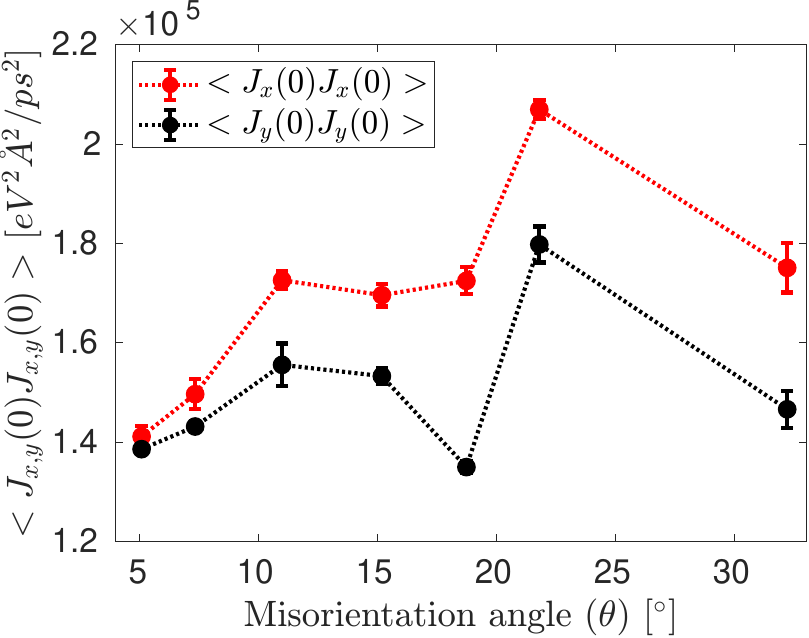} \label{fig17c}}
     \subfloat[$\kappa$ and  $\kappa^{\rm{derived}}$]{\includegraphics[width=0.45\textwidth]{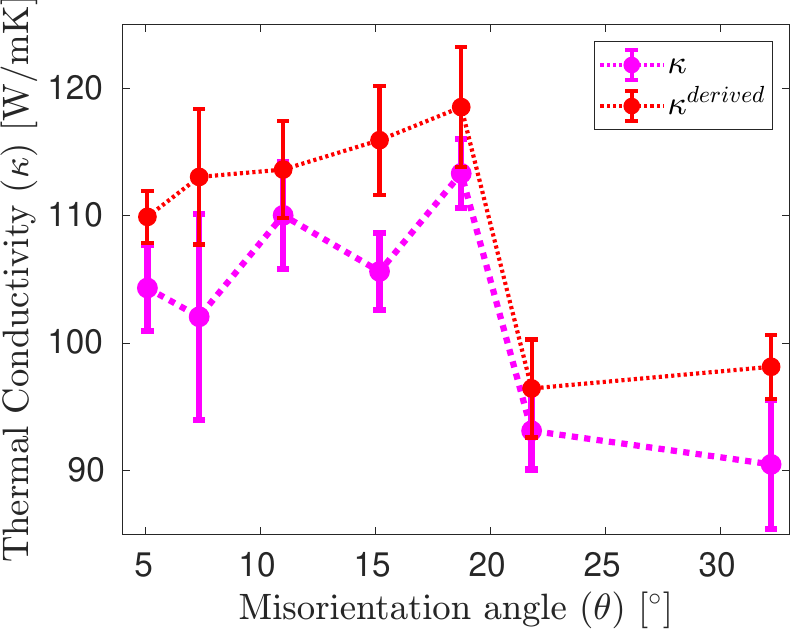} \label{fig17d}}
     \caption{For different tilt angles, values of (a) 
    $\kappa_{\rm{xx}}^{\rm{derived}}$ and $\kappa_{\rm{yy}}^{\rm{derived}}$ in W/mK,
    (b) $A_{1_x}\tau_{1_x}+A_{2_x}\tau_{2_x}$ and $A_{1_y}\tau_{1_y}+A_{2_y}\tau_{2_y}$ in ps, 
    (c) $\avg{J_x(0)J_x(0)}$ and $\avg{J_y(0)J_y(0)}$ in eV$^2$\AA$^2$/ps$^2$,
    and (d) $\kappa$ and  $\kappa^{\rm{derived}}$ in W/mK.
     The red color in (a)-(c) represents the x-component values and the black color
     represents the y-component values.}
    \label{fig17}
\end{figure}
Similarly to our analysis of the curvature in \sect{sec:curvature}, we plot $\kappa_{\rm{xx}}^{\rm{derived}}$ and $\kappa_{\rm{yy}}^{\rm{derived}}$ in \fig{fig17a}, $A_{1_{c}}\tau_{1_{c}}+A_{2_{c}}\tau_{2_{c}}$ in \fig{fig17b}
and $\avg{J_c(0)J_c(0)}$ in \fig{fig17c},  where $c=\rm{x}$ or $\rm{y}$.
Only $A_{1_{c}}\tau_{1_{c}}+A_{2_{c}}\tau_{2_{c}}$ follows the same trends with
respect to tilt angles as $\kappa_{\rm{cc}}^{\rm{derived}}$. 
We also plot $\kappa^{\rm{derived}}$ for different tilt angles in \fig{fig17d} and
find that they are in agreement with the values of $\kappa$. This means that the two measures of the TCs are equivalent to each other.

\section{Conclusion}\label{sec:conclusion}
In conclusion, we study the impact of ripples, curvature, and grain boundary tilt angles on the TCs of SLG. With the help of Green-Kubo simulations on
larger samples and SED simulations on smaller samples, we were able to
decouple the effects of ripples and tensile strain on the TCs of SLG.
Both ripples and tensile strain decrease the TC of the pristine graphene sample.
We show that between 0 and 3\% strain, the strain alone reduces the TC by around
36\%, and between 3\% and 10\% strain, the strain is responsible for a reduction of around 30\% in the TC. We also show that ripples alone can reduce the TC of larger unstrained samples by approximately 61\%. We also observe approximately 46\% decrease in $\rm{v_{{g}_{rms}}^2}$ when the strain increases from 0 to 10\%. However, the mean lifetime fluctuates, but the overall magnitude appears to decrease for higher strain values.
We also introduced curvatures in the graphene sheet containing GBs under specific pressure conditions with NPT ensembles and then characterized them with principal curvatures. We reveal that the Green-Kubo TC linearly decreases with maximum curvature, and the rate of decrease depends on the tilt angles of the grain boundary. We also calculated relaxation times by fitting double exponential decaying functions to the normalized HCACFs for different curvatures and showed some interesting trends. We also showed strong anisotropy in the TC values for flat or curved graphene samples with grain boundaries. The TC values also strongly depend on the tilt angles. Due to time constraints, we were unable to perform the ripples and curvature analysis for samples of many different sizes and grain boundaries with all possible tilt angles at different temperatures. This work can be pursued in the future, and even a better understanding of ripples, curvature, and tilt angles can be developed for single- or multiple-layer graphene systems.

\begin{acknowledgments}
This work was supported by the Department of Science \& Technology, India under the SERB grant MTR/2021/000550. The authors appreciate the financial support and computing resources of IIT Bombay. The views expressed in this paper are those of the authors and neither of their affiliated institutions nor of the funding agencies.
\end{acknowledgments}

\bibliography{main}

\end{document}